\documentclass[12pt,preprint]{aastex}
\usepackage{emulateapj5,apjfonts}

\shorttitle{Orosz et al.}
\shortauthors{A Black Hole in XTE J1550-564}

\begin{document}

\title{Dynamical Evidence for a Black Hole in the Microquasar XTE 
J1550-564$^1$}

\author{Jerome A. Orosz}
\affil{Astronomical Institute, Utrecht University, Postbus 80000,
3508 TA Utrecht, The Netherlands}
\email{J.A.Orosz@astro.uu.nl}

\author{Paul J. Groot}
\affil{Harvard-Smithsonian Center for Astrophysics, 60 Garden Street,
Cambridge, MA 02138}
\email{pgroot@cfa.harvard.edu}

\author{Michiel van der Klis}
\affil{Astronomical Institute ``Anton Pannekoek,'' University of Amsterdam
and Center for High-Energy Astrophysics, Kruislaan 403, 1098 SJ
Amsterdam, The Netherlands}
\email{michiel@astro.uva.nl}

\vspace{1ex}

\author{Jeffrey E. McClintock, Michael R. Garcia, and Ping
Zhao}
\affil{Harvard-Smithsonian Center for Astrophysics, 60 Garden Street,
Cambridge, MA 02138}
\email{jem,garcia,zhao@cfa.harvard.edu}

\author{Raj K. Jain}
\affil{Department of Physics, Yale University, P.O. Box 208120, New Haven,
CT 06520-8120}
\email{rjain@astro.yale.edu}

\author{Charles D. Bailyn}
\affil{Yale University, Department of Astronomy, P.O. Box 208101,
New Haven, CT 06520-8101}
\email{bailyn@astro.yale.edu}

\and

\author{Ronald A. Remillard}               
\affil{Center for Space Research, Massachusetts Institute of Technology,
Cambridge, MA 02139-4307}
\email{rr@space.mit.edu}

\altaffiltext{1}{Based on observations collected at the 
European Southern Observatory, Chile (program 67.D-0229),
at the Magellan Walter Baade telescope
at the Las Campanas 
Observatory, Chile,
and the William Herschel Telescope 
operated on the island of La Palma by the Isaac Newton 
Group in the Spanish Observatorio del Roque de los
Muchachos of the Instituto de Astrofisica de Canarias}

\begin{abstract}
Optical spectroscopic observations of the companion star (type G8IV to
K4III) in the microquasar system XTE J1550-564 reveal a radial
velocity curve with a best fitting spectroscopic period of $P_{\rm
sp}=1.552\pm 0.010$ days and a semiamplitude of $K_2=349\pm 12$ km
s$^{-1}$.  The optical mass function is $f(M)=6.86\pm 0.71\,M_{\odot}$
($1\sigma$).  We tentatively measure the rotational velocity of the
companion star to be $V_{\rm rot}\sin i=90\pm 10$ km s$^{-1}$, which
when taken at face value implies a mass ratio of $Q\equiv
M_1/M_2=6.6^{+2.5}_{-1.6}$ ($1\sigma$), using the above value of
$K_2$.  We derive constraints on the binary parameters from
simultaneous modelling of the ellipsoidal light and radial velocity
curves.  We find $1\sigma$ ranges for the photometric period
($1.5430\,{\rm d}\le P_{\rm ph}\le 1.5440\,{\rm d}$), $K-$velocity
($350.2\le K_2\le 368.6$ km s$^{-1}$), inclination ($67.0^{\circ}\le
i\le 77.4^{\circ}$), mass ratio ($Q\ge 12.0$), and orbital separation
($11.55\,R_{\odot}\le a\le 12.50\,R_{\odot}$).  Given these
geometrical constraints we find the most likely value of the mass of
the compact object is $9.41\,M_{\odot}$ with a $1\sigma$ range of
$8.36\,M_{\odot}\le M_1\le 10.76\,M_{\odot}$.  If we apply our
tentative value of $V_{\rm rot}\sin i=90\pm 10$ km s$^{-1}$ as an
additional constraint in the ellipsoidal modelling, we find $1\sigma$
ranges of $1.5432\,{\rm d}\le P_{\rm ph}\le 1.5441\,{\rm d}$ for the
photometric period, $352.2\le K_2\le 370.1$ km s$^{-1}$ for the
$K-$velocity, $70.8^{\circ}\le i\le 75.4^{\circ}$ for the inclination,
$6.7\le Q\le 11.0$ for the mass ratio, and $12.35\,R_{\odot}\le a\le
13.22\,R_{\odot}$ for the orbital separation.  These geometrical
constraints imply the most likely value of the mass of the compact
object of $10.56\,M_{\odot}$ with a $1\sigma$ range of
$9.68\,M_{\odot}\le M_1\le 11.58\,M_{\odot}$.  In either case the mass
of the compact object is well above the maximum mass of a stable
neutron star, and we therefore conclude XTE J1550-564 contains a black
hole.
\end{abstract}


\section{Introduction}

X-ray novae provide the strongest evidence for the existence
of stellar mass black holes.  These objects are interacting binaries
containing a compact primary (a neutron star or a black hole) and what
is usually a late-type secondary.  X-ray novae spend most of
their time in a low X-ray luminosity ``quiescent'' state, 
with an X-ray luminosity ($L_x$) roughly similar to the optical luminosity
($L_{\rm opt}$), which includes a substantial contribution from
the secondary star ($L_{\rm star}$).
The quiescent state is occasionally interrupted by
``outbursts'' where typically
$L_x\gg L_{\rm opt}\gg L_{\rm star}$.  In quiescence, the observed
radial velocity and light curves of the secondary star lead to
dynamical mass estimates for the compact primary. If its mass exceeds
the maximum stable mass of a neutron star ($\approx 3\,M_{\odot}$), 
the compact object
is presumed to be a black hole \citep{chi76,kal96}.  In thirteen cases, the
mass of the primary of an X-ray nova has been shown to exceed
$3\,M_{\odot}$, confirming the presence of black holes in these
systems \citep{bai95,cas92,cas95,fil95a,fil99,fil01,gre01,
mcc86,mcc01,oro98,oro01,rem92, rem96}.
These X-ray novae 
open up the possibility of studying the
strong-field regime
of general relativity.  For example, quiescent X-ray spectra and ADAF
(advection-dominated accretion flow) models provide evidence for black
hole event horizons 
\citep{nar96,gar01}. Also,
the study of high-frequency QPOs (quasi-periodic
oscillations) may lead to the first secure measurement of black hole
spin (e.g.\ Remillard et al.\ 1999; Strohmayer 2001).

XTE J1550-564 was discovered on 1998 September 7
by the All-Sky Monitor (ASM) aboard the
{\em Rossi X-ray Timing Explorer (RXTE)} 
\citep{smi98}. 
The optical counterpart (designated V381 Normae) and the radio
counterpart  were  discovered shortly thereafter \citep{oro98a,cam98}.
This source was quickly identified as a promising
black hole candidate based on its rapid X-ray variability, hard energy
spectrum, and the absence of pulsations or X-ray bursts
\citep{cui99,sob99}.  Relativistic plasma ejections at probable
superluminal velocities were observed at radio wavelengths shortly
after the strong X-ray flare in 1998 September 
\citep{han01b},
indicating that XTE J1550-564 is another microquasar.  The galactic
microquasars are excellent laboratories for the study of relativistic
jets since they evolve orders of magnitude more quickly than do the
jets in quasars \citep{mir99}.  XTE J1550-564 is also of special
interest owing to its complex X-ray variability (e.g.\ Homan et al.\
2001;
Remillard et al.\ 2002).  In this paper we report the results of our 
recent optical 
observations of XTE J1550-564.  The observations and basic data
reductions are summarized in  \S\ref{obssec}.  In 
\S\ref{ansec} we establish the orbital parameters of the system,
derive some properties of the secondary star, derive geometrical
parameters for the binary, and discuss limits
on the mass of the compact object.  The implications of our results
are discussed in \S\ref{dissec} and summarized
in \S\ref{sumsec}.

\section{The Observations and Their Reductions}\label{obssec}

\subsection{Spectroscopy}

We obtained a total of 18
spectra of the source on 2001 May 24-27 
using Antu, which is the first 8.2 m
telescope at the European Southern Observatory, Paranal.  
We used the FORS1 imaging spectrograph with
the
600V grism  and a 0\farcs 7 wide
slit;
this combination gives a spectral resolution of 3.6~\AA\ FWHM and
a wavelength coverage of 4571-6927~\AA.  
The seeing was better than 0\farcs 8 on May 24 and May 27,
and relatively poor on May 26 ($> 1\farcs 5$). 
Clouds were present on the night of May 25, and only one useful spectrum
was obtained in 1\farcs 5 seeing.  
The
exposure times ranged from 20 to 40 minutes depending on the conditions.
We also observed eight bright subgiant (luminosity class IV)
and giant (luminosity class III) stars with spectral types from G8 to M0.
An atmospheric dispersion corrector was used and the slit was kept at
the default north-south direction.  
Following the standard procedure
at Paranal, the flat-field and wavelength calibration exposures were
obtained during the daytime hours with the telescope pointed at the
zenith.

We reduced all of the spectra using 
IRAF\footnote{IRAF is distributed by the National Optical Astronomy 
Observatories,
which are operated by the Association of Universities for Research
in Astronomy, Inc., under cooperative agreement with the National
Science Foundation.}.  The standard tasks were
used to apply the bias and flat-field corrections, and to extract and
wavelength-calibrate the spectra.  The night-sky emission lines
were used to make small adjustments to the
wavelength scales of the spectra.  
The spectra were placed on an approximate flux scale using
observations (taken May 27) of the
white dwarf flux standards EG 274 and Feige 110.  
The final reduced spectra of XTE J1550-564 have typical 
signal-to-noise ratios in the continuum near 6 per 1.16~\AA\
pixel
at 6000~\AA\ and about 2 per 1.16~\AA\
pixel at 5000~\AA.

We extended our grid of  comparison stars by making use
of archival 
observations.  
Eleven spectra of seven different stars were taken from
the FORS1 archive\footnote{ESO programs 63.N-0481,
65.H-0360,
and 265.D-5016}.
These spectra were obtained using the
600R grism and a 0\farcs 7 wide
slit;
this combination gives a spectral resolution of 3.2~\AA\ FWHM and
a wavelength coverage of 5142-7283~\AA.
All of the standard reductions were performed with IRAF.
We also used 49 spectra of 34 different stars obtained 
1995 April 30 and 1995 May 2-4 using the 
CTIO 4m telescope and the RC spectrograph and the Loral
$3072\times 1024$ CCD
(see Bailyn et al.\ 1995; Orosz \& Bailyn 
1997).  The spectral resolution is 3.3~\AA\ 
FWHM and the wavelength coverage is 3850-7149~\AA.
Finally, we used moderate resolution 
spectra of 6 stars with spectral types K5V to M0V obtained 
on 2001 February 16 and 17 with the
red arm of the ISIS instrument on the 4.2m William Herschel Telescope
(WHT) on La Palma.  The combination of the 600R grating with the
$1024\times 1024$ TEK4 CCD and a 1 arcsecond slit width yields a
resolution of $\approx 70$ km s$^{-1}$ FWHM and a wavelength coverage
of 5856-6646~\AA.
These spectra were reduced with IRAF.

\subsection{Photometry}

We observed XTE J1550-564 2001 June 26-28 using the 
Magellan Instant Camera (MagIC) on the 
Walter Baade
6.5m telescope located at Las Campanas Observatory.
The MagIC contains a SITe $2048\times 2048$ CCD with a scale of 0\farcs 069
per pixel.  The source was observed for about 5 hours on each of the three
nights in generally good conditions (0\farcs 3 to 0\farcs 7 seeing and thin
cirrus) with the Sloan $r^{\prime}$, $i^{\prime}$, and $z^{\prime}$ filters.
Some images in the Sloan $g^{\prime}$ filter were also obtained on
the night of June 26.

XTE J1550-564 has been monitored extensively by the 1m YALO
telescope (Bailyn et al.\ 2000)
at Cerro Tololo Interamerican Observatory 
since  the discovery of the optical counterpart in 1998 September 
\citep{oro98a}.  The $V$-band data from 2001 discussed here were collected and
reduced in a similar manner to previous data from this source 
described by \citet{jai01b,jai01c}.

As part of the normal acquisition procedure for spectroscopy
we obtained 
a total of
nine direct $V$ images with exposure times of
30 to 90 seconds with VLT/FORS1 in its  imaging 
mode.  In addition, 
two additional
direct images in $B$ with exposure times of 120 and
200 seconds were taken.
The image scale of FORS1 is 0\farcs 2 per pixel, and the field of
view is $6.8\times 6.8$ square arcminutes.

Finally, photometry of XTE J1550-564 was obtained 2001 June 1 using the
SuSI2 instrument on the Nasmyth  focus of the
3.5m New Technology Telescope (NTT) located at  
the European Southern Observatory, 
La Silla.   SuSI2
is a mosaic of two $2048 \times 4096$, thinned, anti-reflection coated
EEV CCDs.  We used $3\times 3$ on-chip binning, yielding a scale
of 0\farcs 24 per pixel and a field of view of $5.5 \times 5.5$ square
arcminutes.  The observing conditions were quite poor: the seeing was between
1\farcs 5 and 2\farcs 0 and there were passing clouds.
We obtained
a total of 51 usable $R$-band images and 4 usable $V$-band images with 
exposure times of 3 to 5 minutes.

The image processing routines
in IRAF  were used to correct
for the electronic bias and to apply the flat-field corrections to the VLT,
NTT, and Magellan images.
The programs {\sc DAOPHOT IIe}, {\sc ALLSTAR} and
{\sc DAOMASTER} 
(Stetson 1987; 
Stetson, Davis, \& Crabtree 1991; 
Stetson 1992a,b) were used
to compute the instrumental magnitudes
of XTE J1550-564 and all of the
field stars within about
a 1.2 arcminute radius.  
The comparison stars ``A'', ``B'', and ``C'' shown in
Figure 2 of \citet{jai01b} were used to place the 
$B$- and $V$-band instrumental magnitudes
on the standard  scales.  
The $R$-band photometry and the Magellan photometry were left on the
instrumental magnitude scales.  Since some of the $V$-band
images from the VLT and NTT were slightly under-exposed, we averaged
groups of two to four consecutive exposures together, yielding two 
measurements from 24 May, and one measurement each from the nights
of 25-27 May  and 1 June.

\section{Analysis}\label{ansec}

\subsection{Long-term Photometry}

\begin{figure}
\epsscale{0.7}
\plotone{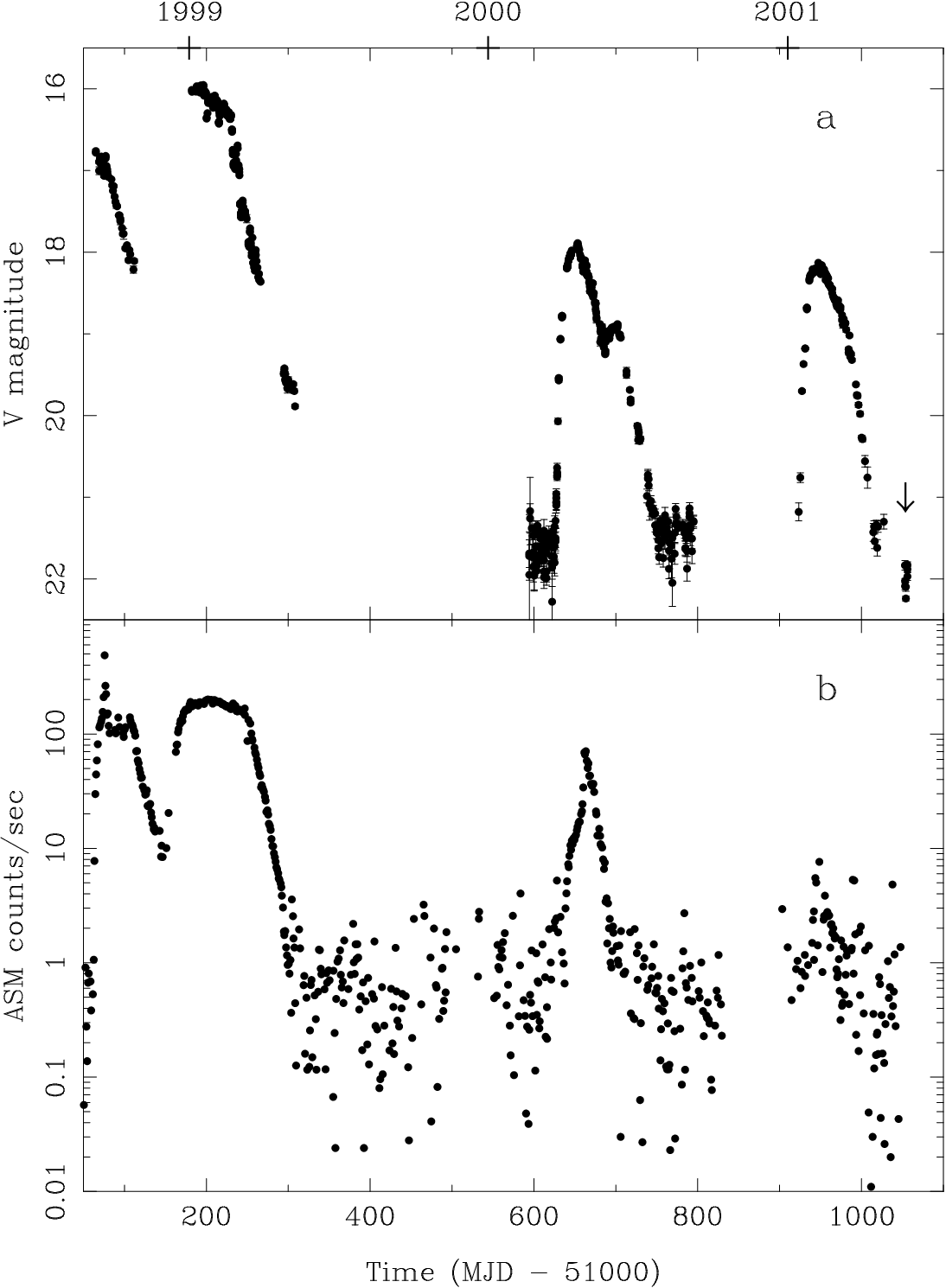}
\figcaption[f01.eps]{(a): 
The complete YALO $V$-band light curve of XTE J1550-564 is shown to the
left of the arrow located near day 1050.  The arrow indicates the time
of our spectroscopic observations and our $V$-band photometric observations
with VLT/FORS1, which are shown here.
(b) The {\em RXTE} 
ASM 2-12 keV light curve 
of XTE J1550-564
(daily averages).  The $y$-axis scale is logarithmic to highlight the
weaker activity in early 2001 (the measurements with negative flux are not
shown).
\label{fig0}}
\end{figure}

Figure \ref{fig0}a shows the complete YALO 
$V$-band light curve together with our VLT photometry. 
The 
2-12 keV X-ray light curve from the All Sky Monitor (ASM) on
{\em RXTE} is shown in Figure \ref{fig0}b.    
XTE J1550-564 had an optical reflare in 2001 January
\citep[Fig.\ \protect\ref{fig0}a]{jai01a}.  Strangely enough, the 
corresponding outburst in X-rays was quite weak, with a peak level of
less than 10 ASM counts per second.  
The source returned to its 
quiescent level by about April 20, and our $V$-band magnitudes
from May 24-27 are  fully consistent with the quiescent level:
$21.83\le V \le 22.24$, with typical errors of 0.04 mag.

The source was not detected in $B$, and we place a conservative lower
limit of $B>24.0\pm 0.1$.  \citet{jai99} had estimated $B=22.0\pm 0.5$
for the quiescent level  based on a detection (near the plate limit)
of the source on the SERC J survey print.  The effective bandpass of the
SERC J print is actually much redder than the standard Johnson $B$ 
filter\footnote{http://www.roe.ac.uk/ukstu/platelib.html\#eqj},
so the quiescent magnitude given in \citet{jai99} is not a proper $B$
magnitude but is closer to a $V$ magnitude.

\subsection{Orbital Period and Spectroscopic Elements}

We measured the radial velocities of the secondary star using the {\em fxcor}
task within IRAF, which is an implementation of the technique of
\citet{ton79}.  The cross correlations were computed over the wavelength
range 5000-6850~\AA, excluding the H$\alpha$ emission line, interstellar
lines (Na D and the diffuse band near 5876~\AA), and a telluric feature
near 6280~\AA.  The spectra were continuum-subtracted prior to computing
the cross correlation, and a Fourier filter was
used to remove high frequency noise.  
With  the exception of one spectrum from the end of the night
of May 27 which had poor signal-to-noise,
the cross correlation peaks were generally quite significant. 
The value of the Tonry \& Davis ``$r$'' parameter, which is a measure
of the signal-to-noise in a cross correlation, was generally in
the range of 3.0-4.8.
The spectrum of the K3III star HD 181110 usually
gave the best cross correlation
peaks as judged
by the value of the $r$ parameter, 
and the spectrum of the K4III star HD 181480 generally gave the
second-best cross correlation peaks.  In the analysis below
we adopt the velocities measured using the K3III template.
Since the source was at or very near quiescence,
we believe our radial velocities are not
biased by X-ray heating.

\begin{figure}
\epsscale{0.7}
\plotone{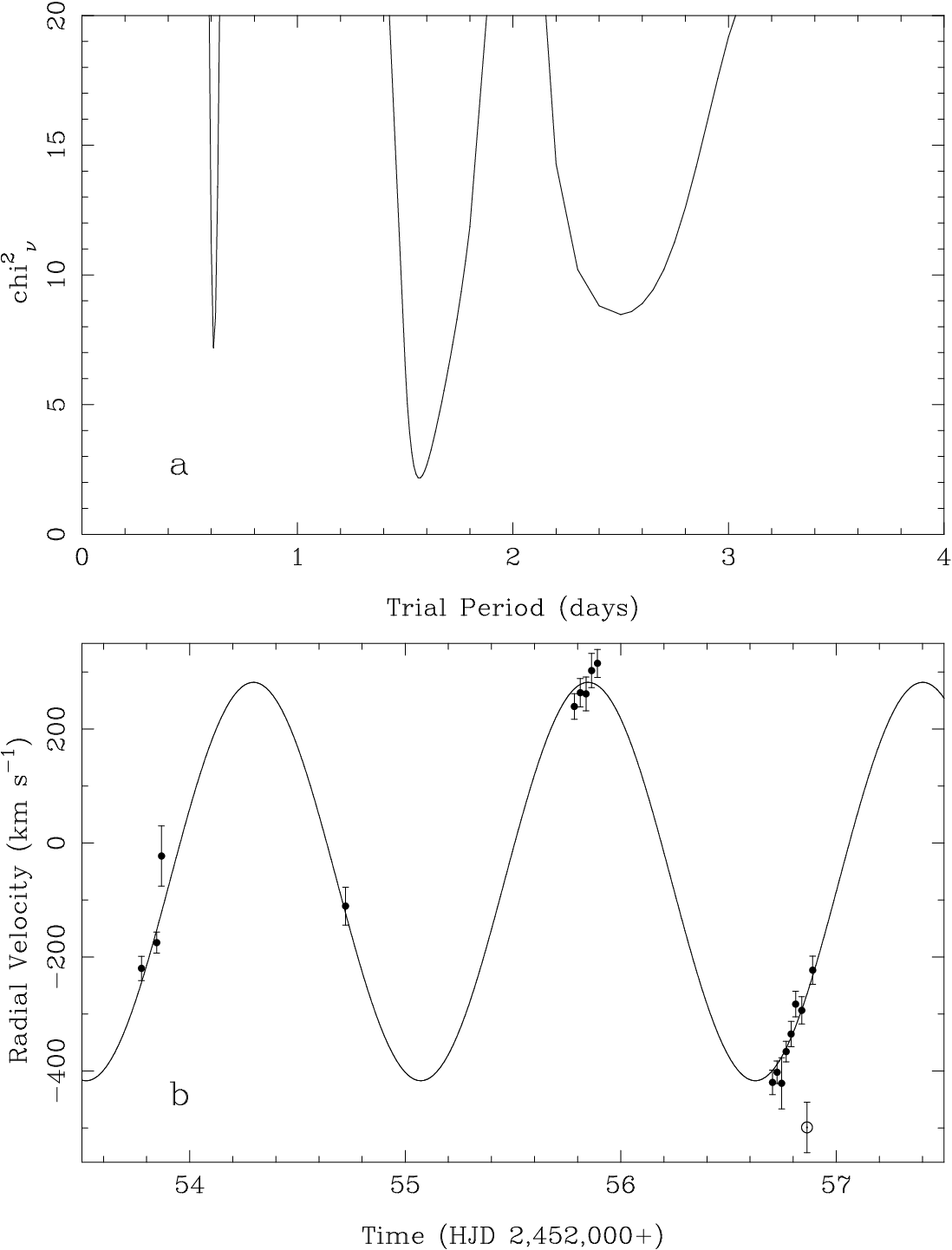}
\figcaption[f02.eps]{(a): The reduced
$\chi^2_{\nu}$ for a three-parameter sinusoidal fit to the 17 radial
velocities as a function of the trial period in days.  The minimum
$\chi^2_{\nu}$ is at $P=1.552\pm 0.010$ days, and the typical $\chi^2_{\nu}$
value is 40.
(b): The radial velocities and the best fitting
sinusoid, plotted as a function of time.
The velocity plotted as a open circle has been excluded from the fit.
\label{fig1}}
\end{figure}

To search for the spectroscopic period we computed a three-parameter
sinusoidal fit to the 17 velocities for a dense
range of trial periods between 0
and 4 days.
The reduced $\chi^2$ values for these fits are shown in Fig.\ 
\ref{fig1}a.
The free parameters at each trial period are the velocity
semiamplitude $K_2$, the epoch of maximum velocity $T_0$(spect), and the
systemic velocity $\gamma$. 
The best fit is at a period
of $P_{\rm sp}=1.552\pm 0.010$ days ($1\sigma$ error),
where $\chi^2_{\nu}=1.38$.
The two alias periods near 0.7 and 2.5 days are 
ruled out by their large values of $\chi^2_{\nu}$ and by
inspection of the folded velocity curves. We
adopt the following spectroscopic elements
with $1\sigma$ errors: $P_{\rm sp}=1.552\pm
0.010$ days, $K_2=349\pm 12$ km s$^{-1}$, and $T_0({\rm
spect})={\rm HJD~}2,452,054.296\pm 0.014$.
In order to find the true systemic velocity we must know the radial
velocity of the template star.  We cross correlated the spectrum
of HD 181110 with 47 template spectra of stars with a radial velocity
measurement
listed in the SIMBAD database.  The median heliocentric velocity 
for HD 181110 was $-71$ km s$^{-1}$ and the standard deviation of
the velocities was 18 km s$^{-1}$.  
Based on this, we adopt   
$\gamma=-68\pm 19$ 
km s$^{-1}$
for XTE J1550-564.
The velocities and the best fitting sinusoid are shown in Fig.\ 
\ref{fig1}b,
and the spectroscopic elements are listed in Table \ref{tab2}.
(The velocity plotted as an open circle in Fig.\ 
\ref{fig1}b
was
excluded from the fit since it deviates by more than $4\sigma$
from the fit.)
We note that our 
spectroscopic period is consistent with the photometric
period of $P_{\rm ph}=1.541\pm 0.009$ days found by \citet{jai01b}.
The optical mass
function
is then 
\begin{equation}
f(M) \equiv {P_{\rm sp}K^3_2\over 2\pi G} = {M_1^3\sin^3i\over (M_1+M_2)^2}
= 6.86\pm 0.71\,M_{\odot}.
\label{fm}
\end{equation}
($1\sigma$ error).  
The optical mass function 
places an immediate lower limit on the 
mass of the compact object, and since
this minimum mass of the compact object  is well above
the maximum mass of a stable neutron star ($\approx 3\,M_{\odot}$,
Chitre \& Hartle 1976; Kalogera \& Baym 1996),  
we conclude XTE J1550-564 contains a black hole.
See \S\ref{inclcon} for further discussion
of the mass of the black hole.

\subsection{Parameters for the Secondary Star}\label{secsec}

Using the spectroscopic
ephemeris in Table \ref{tab2}, the eleven spectra with
the highest signal-to-noise were Doppler corrected
to zero velocity and averaged to create a
``restframe'' spectrum.  The restframe spectrum was dereddened using
the extinction law of \citet{car89}, 
where
we adopted a visual extinction of $A_V=4.75$ magnitudes
[$N_H=(8.5^{+2.2}_{-2.4})\times 10^{21}$ cm$^{-2}$, Tomsick et al.\
(2001) and $A_V=N_H/1.79\times 10^{21}$, \citet{pre95}].
The true extinction is unlikely to be greater than 4.75 mag, since
the extinction derived from X-ray measurements of $N_H$
nearly always exceeds the optically measured value
\citep{vrt91}.

\subsubsection{Spectral Type and Effective Temperature}

\begin{figure}
\epsscale{0.7}
\plotone{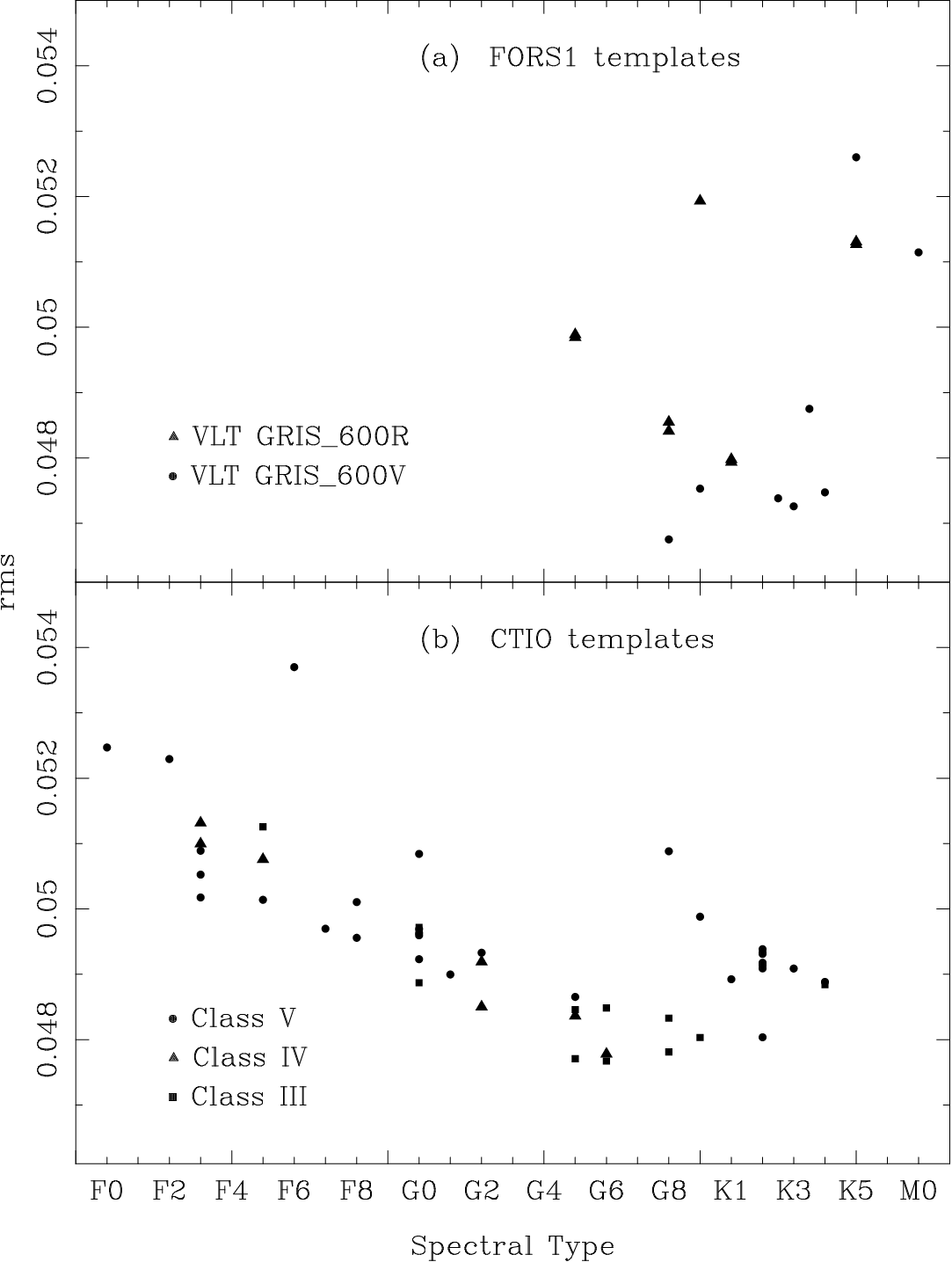}
\figcaption[f03.eps]{(a):  The rms values of the polynomial
fits to the difference spectra as a function of the template spectral
type for the templates obtained with VLT/FORS1 (luminosity class III
and IV).
The plotting symbols
indicate the grism used.
(b):
Same as (a) but for the templates observed at CTIO.  Here the plotting
symbols indicate the luminosity class of the template.
\label{plotnewrms}}
\end{figure}

We used the technique outlined in \citet{mar94} to decompose the restframe
spectrum into its disk and stellar components.
For this purpose we have three sets of template spectra:  eight spectra
of eight different stars taken with the FORS1 600V grism
(resolution $\sim 3.6$~\AA), eleven spectra of seven different stars taken
with the FORS1 600R grism (resolution $\sim 3.6$~\AA),
and 49 spectra of 34 different stars taken at CTIO (resolution
$\sim 3.3$~\AA).
The dereddened restframe spectrum and the template spectra
were all normalized to unity at 6000~\AA\
and resampled to a pixel size of 
1.155~\AA.  
The template spectra were then scaled by various values
of a weight factor $w$ between 0.0 and 1.0 in steps of 0.02, and
subtracted from the restframe spectrum.  The scatter in each
difference spectrum was measured by fitting a low-order polynomial and
computing the rms difference over the same wavelength region used for
the cross correlation analysis above.
For each template spectrum we looked for the value of 
$w$ that gave
the ``smoothest'' difference spectrum
(i.e.\ the $w$ with the lowest associated rms).  
Figure \ref{plotnewrms} shows the rms values as a function of the template
spectral type for the three sets of templates.  Since the spectral resolution
in each of the three sets is slightly different, one should only compare
the rms values within a single set.
For the CTIO grid
the minimum rms is near spectral type G6.  However, the difference
spectra for the G6 templates (and the templates of earlier type) have
negative flux in the blue parts of the spectrum, which is unphysical.
This difficulty could be lessened if one were to adopt a larger
value of $A_V$; however, values larger than $A_V=4.75$ mag
are unlikely (see above).  The templates with spectral type G8 and later,
on the other hand, all produce subtracted spectra with positive differences.
The minimum rms values for the FORS1 
templates is at spectral type G8 for the grism 600V and spectral type K1
for the grism 600R.   In general, there is a steep rise in the rms 
values for template spectral types later than K5 and earlier than G5.
The weight factor $w$ is roughly correlated with the template spectral type
where templates with later spectral types have smaller values of $w$.

\begin{figure}
\epsscale{0.7}
%\plotone{f04.eps}
\vspace{250pt}
\includegraphics{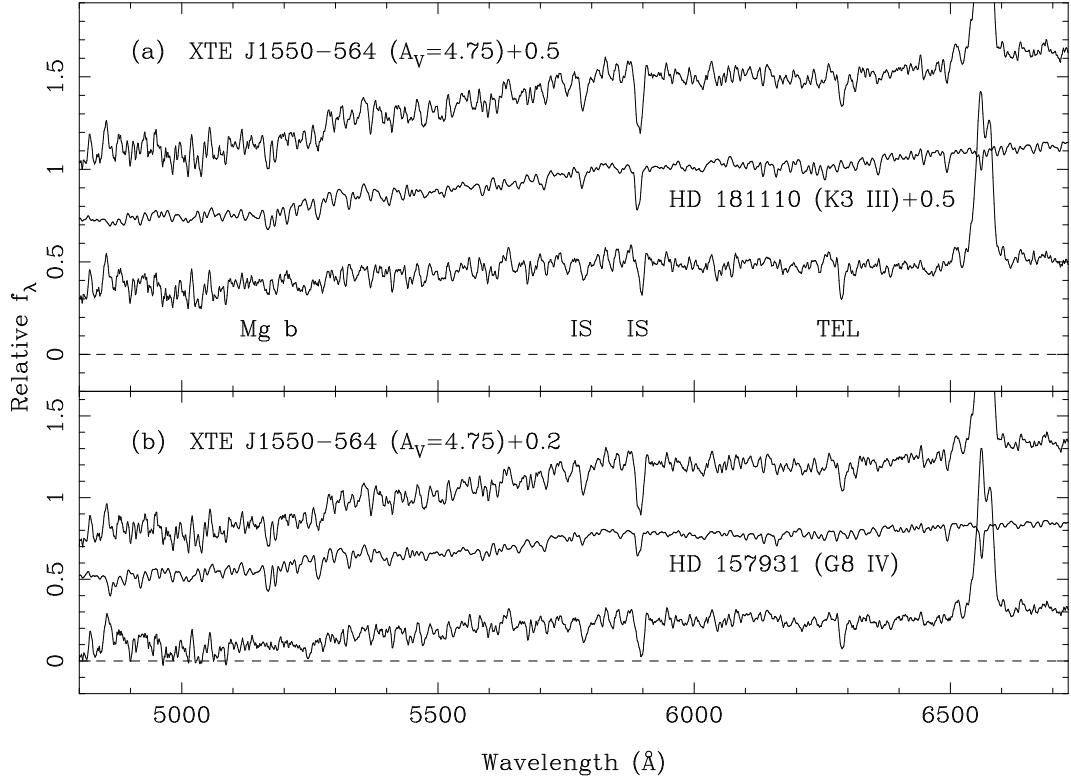}
\figcaption[f04.eps]{The results of the spectral decomposition.
(a)~Top: The dereddened restframe spectrum of XTE J1550-564, smoothed to
5 pixels and offset by 0.5 units.
(a)~Middle:  The spectrum of the K3 III star HD 181110, scaled by 0.48 and
offset by 0.5 units.
(a)~Bottom:  The difference spectrum, which represents the spectrum of the
accretion disk,
smoothed to 5 pixels with no offset.
The interstellar features are denoted by IS and the telluric feature
by TEL.
(b)~Top: The dereddened restframe spectrum of XTE J1550-564, smoothed to
5 pixels and offset by 0.2 units.
(b)~Middle:  The spectrum of the G8IV star HD 157931, scaled by 0.74 with
no offset.
(b)~Bottom:  The difference spectrum
smoothed to 5 pixels with no offset.
\label{fig3}}
\end{figure}

Fig.\ \ref{fig3} shows the restframe spectrum decomposed into the disk and
stellar components for two different templates 
observed with the FORS1 grism 600V (i.e.\ the same instrumental configuration
used for the XTE J1550-564 spectra).  
Fig.\ \ref{fig3}a shows the decomposition using
the K3III template HD 181110 ($w=0.48$) and
Fig.\ \ref{fig3}b shows the decomposition using
the G8IV template HD 157931 ($w=0.74$).
In both cases
the disk spectrum is basically flat in $F_{\lambda}$, and the
only obvious
features remaining are the H$\alpha$ emission line, the interstellar
lines, and the telluric line (note that these features have been
smeared in the process of making the restframe spectrum).  The Mg b lines
near 5169~\AA\ are matched better with the G8IV template.
Apart from Mg b, the two difference spectra look very similar to the eye.

Based on the decomposition analysis above, we adopt a spectral type of G8-K4
for the secondary star.  The spectral type is almost certainly not 
K7 or
later since the TiO bands near 6230 and 6700~\AA\ become apparent
in dwarfs at K7 and later and in giants
at K5 and later, and these bands are not evident in our spectrum.
Spectral types earlier than about G5 are also firmly ruled out
based on the high rms values (Fig.\ \ref{fig3}b) and the partly negative
difference spectra.
The secondary star cannot have a normal main sequence gravity 
owing to the relatively long orbital period, so its surface
gravity must be somewhat less than the nominal  values
for luminosity class V.
The surface gravity of the secondary is easily found using
Kepler's Third Law, assuming that the star fills its Roche lobe.
In that case,
there is a 
relationship between the mean density
of the secondary star and the orbital period:
\begin{equation}
\rho={M_2\over \slantfrac{4}{3}\pi 
R^3_2}={3\pi\over P^2GR^3_{RL}(Q)}{1\over 1+Q}
\label{den}
\end{equation}
where
$R_{RL}(Q)$ is the sphere-equivalent radius of the
Roche lobe for unit separation, and $G$ is the universal constant of
gravity.  One can readily show through numerical integration that
for $2\le Q\le 20$, the quantity $R^{-3}_{RL}(Q)(1+Q)^{-1}$ 
only varies by 4.6\% (between 9.7 and 10.2).  Hence, for a fixed
orbital period $P$, the mean density of the secondary
is nearly independent
of the  mass ratio $Q$.  This in turn means that 
the resulting surface gravity is a very
weak function of the assumed secondary star mass $M_2$
($\log g\propto \slantfrac{1}{3}\log M_2$).  
We find  surface gravities of $\log g=3.51,$ 3.61, and 3.72 
for secondary star
masses of $0.5$, 1.0, and $2.0\,M_{\odot}$, respectively.
These values of $\log g$ imply
a luminosity class between III and V [e.g.\ a K4V star nominally has
$\log g=4.60$ and a K4III star nominally has
$\log g=1.7$, Gray (1992)].   
The temperature corresponding to a given spectral type depends on the
luminosity class, where main sequence stars of a given spectral type
(i.e.\ luminosity class V) are hotter than the corresponding giants
(luminosity class III).  Strai\v{z}ys \& Kuriliene (1981) 
give $T_{\rm eff}\sim 4100$~K for a K4III star and $T_{\rm eff}\sim
5100$~K for a G8IV star.
Thus
we adopt $4100 \le T_{\rm eff}\le 5100$~K for the analysis below.

\subsubsection{Rotational Broadening}\label{broadsubsec}

\begin{figure}
\epsscale{0.6}
\plotone{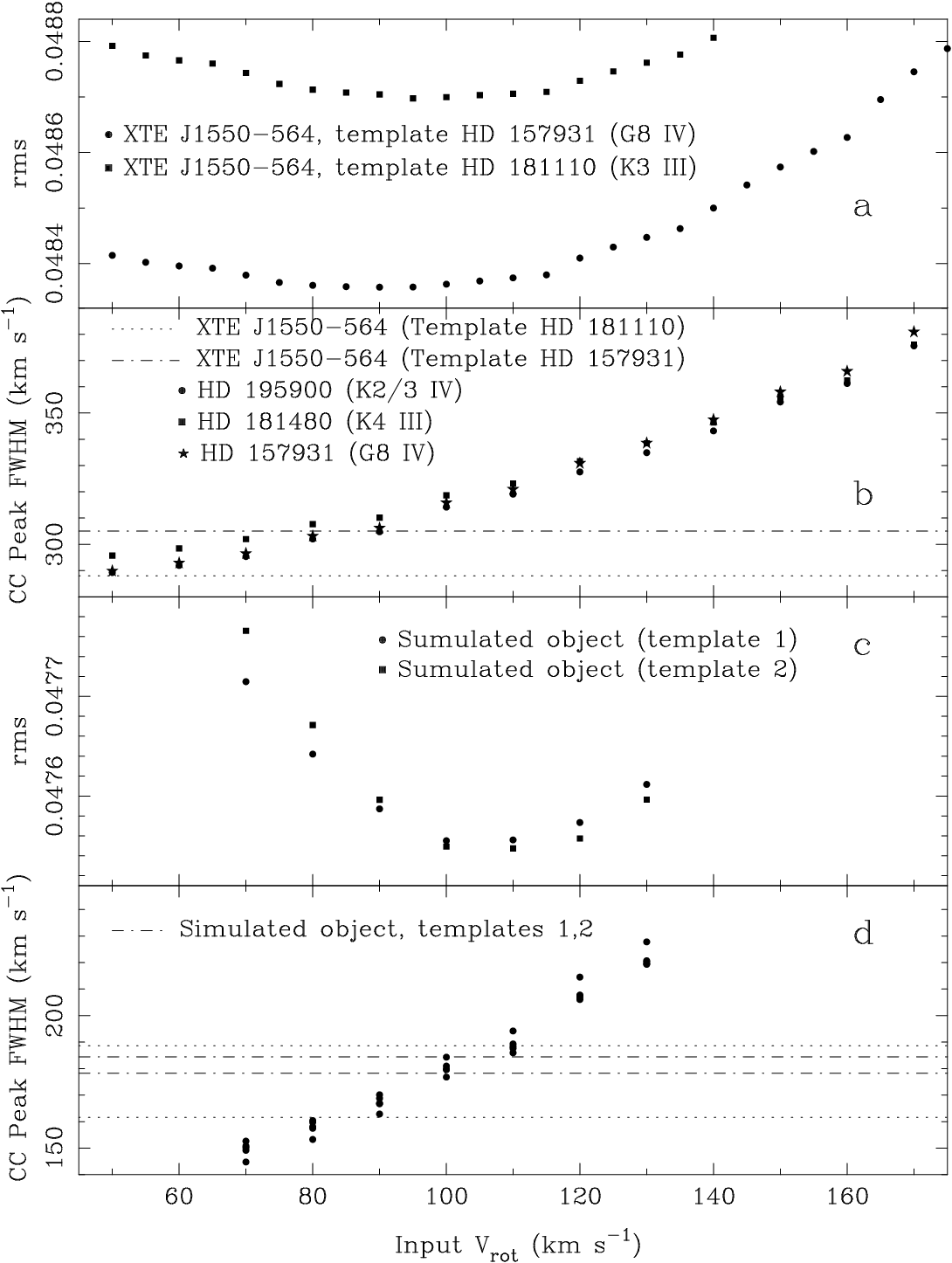}
\figcaption[f05.eps]{(a):  The rms values of the polynomial
fits to the difference spectra 
as a function of the input value of the rotational velocity
$V_{\rm rot}\sin i$.  The lowest rms value is for
$V_{\rm rot}\sin i=90$ km s$^{-1}$ for
the G8IV  template HD 157931 (filled circles) and
$V_{\rm rot}\sin i=95$ km s$^{-1}$ for the K3III template
HD 181110 (filled squares).
(b):
The full width at half maximum of the cross correlation peaks of
three different comparison star spectra as a function of the input
value of the rotational velocity $V_{\rm rot}\sin i$.  In all cases
the template spectrum for the cross correlation analysis was the K3III
star HD 181110.  The  lines show the width of the cross
correlation peak for the XTE J1550-564 restframe spectrum measured using
two different template spectra (the dotted line
for the K3III template HD 181110  and the dash-dotted line for the G8 IV
template HD 157931).  The results in (a) and
(b)
show that the rotational
broadening of XTE J1550-564
is on the order of 90 km s$^{-1}$ or less.
(c):  Similar to (a) but for a simulated spectrum constructed with
a moderate resolution spectrum ($\approx 70$ km s$^{-1}$ FWHM), using
an input value of $V_{\rm rot}\sin i=90$ km s$^{-1}$.
(d):  Similar to (c) but for the simulated spectrum and the 
moderate resolution templates.
The dotted lines show the full range of the width of the cross
correlation peak for the simulated object spectrum measured with
all six templates, and the dash-dotted lines show the width of the
cross correlation peak measured with the two best matched templates. 
The results in (c) and (d) show that the systematic error involved
in measuring the rotational velocity in a relatively noisy low resolution
spectrum is not too large and is perhaps on the order of about 10
km s$^{-1}$.
\label{plotcc}}
\end{figure}

We used the Marsh et al.\ (1994) technique to estimate the 
projected rotational velocity $V_{\rm rot}\sin i$ of the XTE
J1550-564 restframe spectrum.  The dereddened restframe spectrum, 
the spectrum of the G8IV star HD 157931,  and
the spectrum of the K3III star HD 181110
were each normalized to
unity at 6000~\AA\ and resampled to a pixel size of 0.46~\AA.
An IRAF procedure implementing the analytic broadening kernel given in
Gray (1992, p. 374) was used to broaden the
template spectrum using
values of $V_{\rm rot}\sin i$ 
between 50 and 160 km s$^{-1}$ in steps
of 5 km s$^{-1}$ (we adopted a linear limb darkening coefficient of
0.6).  The value of the weight factor $w$ was optimized to find the
minimum rms value at each input value of $V_{\rm rot}\sin i$.
The results are shown in Fig.\ \ref{plotcc}a.  The minimum rms value
occurs for $V_{\rm rot}\sin i=90$ km s$^{-1}$ for the G8IV template
and at $V_{\rm rot}\sin i=95$ km s$^{-1}$
for the K3III template.

We next attempted to measure $V_{\rm rot}\sin i$ for the secondary star using
a different technique that is based on the relationship between the
width of the cross correlation (cc) peak and the rotational velocities
of the object and template spectra \citep{hor86a,ech89,cas93}.
A Gaussian with a fitting width of 750 km s$^{-1}$ and a baseline
of zero was used to determine the width of the cc peaks.  The measured
widths were relatively insensitive to the exact width of the fitting region.
The full width at half maximum (FWHM) of
the restframe spectrum cc peak is in the range of 285 to 305 km s$^{-1}$,
depending on which template is used.  
In order to find out what rotational
velocity this corresponds to, we constructed calibration curves using
some 
bright comparison stars observed with the FORS1 600V grism.
Three different comparison stars were
each broadened by various amounts, and the FWHM of the cc peaks were
measured using the K3III template HD 181110.  The results
are shown in Fig.\ \ref{plotcc}b.  These calibration curves show 
that a cc peak FWHM of 305 km s$^{-1}$ corresponds to a rotational
velocity of $V_{\rm rot}\sin i\approx 90$ km s$^{-1}$, and that a FWHM
of 285 km s$^{-1}$ corresponds to an input $V_{\rm rot}\sin i$ of
about 50 km s$^{-1}$.   

The Marsh et al.\ (1994) technique indicates a rotational velocity of
$V_{\rm rot}\sin i=90$ km s$^{-1}$
(given the better
performance of the G8 template in separating the star
and disk spectra we
adopted the value of $V_{\rm rot}\sin i$ measured with this template, 
Fig.\ \ref{plotcc}a), and the cross correlation technique indicates
$V_{\rm rot}\sin i\lesssim 90$ km s$^{-1}$
(Fig.\ \ref{plotcc}b).
Based on the results of the two techniques, we adopt a value
of $V_{\rm rot}\sin i=90\pm 10$ km s$^{-1}$ for the rotational
velocity of the secondary star.

The value of $V_{\rm rot}\sin i$
should be treated with
caution since our spectral resolution is relatively low (about 160 km
s$^{-1}$) and the signal-to-noise ratio is modest (about 20 per pixel in
the restframe spectrum).  
In order to have some idea of the size of the systematic errors involved
with measuring the rotational velocity in a relatively noisy low resolution
spectrum, we performed some numerical experiments using the moderate
resolution spectra obtained with the WHT.  These experiments are by
no means comprehensive since we have only a limited range of
spectral types (K5 to M0, all dwarfs) and wavelength coverage (roughly
between the Na D lines and H$\alpha$).  The six template spectra were
first normalized at a common wavelength (6450~\AA).
A simulated object spectrum was
made as follows:
a template was scaled by 0.7, rotationally broadened by
90 km s$^{-1}$, degraded to a resolution of 3.6~\AA\ by convolution
with a Gaussian, and resampled to 1.16~\AA\ (to match the pixel size
of the VLT/FORS1 spectra).  Next, a constant value of 0.3 was added
(to simulate the accretion disk), and Gaussian noise was added so that
the signal-to-noise ratio was $\approx 20$.   Finally, the
simulated object spectrum and the remaining template spectra
were resampled to a pixel size of 0.46~\AA.
The rotational velocity of the simulated object
spectrum was extracted by the
two different techniques, and the results are shown in Figs.\
\ref{plotcc}c and d.
The Marsh et al.\ technique gives a rotational velocity of just
over 100 km s$^{-1}$ using the two best matched templates, 
and the cross correlation technique
gives a rotational velocity between about 80 and 110 km
s$^{-1}$ using all templates and between about 100 and 110 km s$^{-1}$
using the two best matched templates.  Given that the input rotational
velocity for the simulated spectrum was 90 km s$^{-1}$, we conclude that
the systematic errors involved in this technique are not excessively large
and are probably on the order of our adopted $1\sigma$ error of 10 km
s$^{-1}$.  Needless to say
it would be desirable to repeat the measurement of the
rotational velocity of the secondary star in XTE J1550-564
using data with better resolution and a higher signal-to-noise
ratio.

If the secondary star is rotating synchronously with the orbit,
its mean rotational velocity depends only on the binary mass ratio 
$Q\equiv M_1/M_2$
and the projected orbital velocity $K_2$
(e.g.\ Horne, Wade, \& Szkody 1986).  We used the ELC code
\citep{oro00} to numerically compute sphere-equivalent Roche lobe radii
as a function of the mass ratio $Q$, and a simple Monte Carlo code
to propagate the ($1\sigma$) uncertainties on the input parameters.  We find
$Q=6.6^{+2.5}_{-1.6}$ ($1\sigma$) for $K_2=349\pm 12$ km s$^{-1}$
and $V_{\rm rot}\sin i=90\pm 10$ km s$^{-1}$.  This mass ratio is relatively
far from unity, which is typical of  X-ray novae with cool
low mass companions (e.g.\ Bailyn et al.\ 1998 and references therein).

\subsection{The H$\alpha$ Profile}

\begin{figure}
\epsscale{0.7}
\plotone{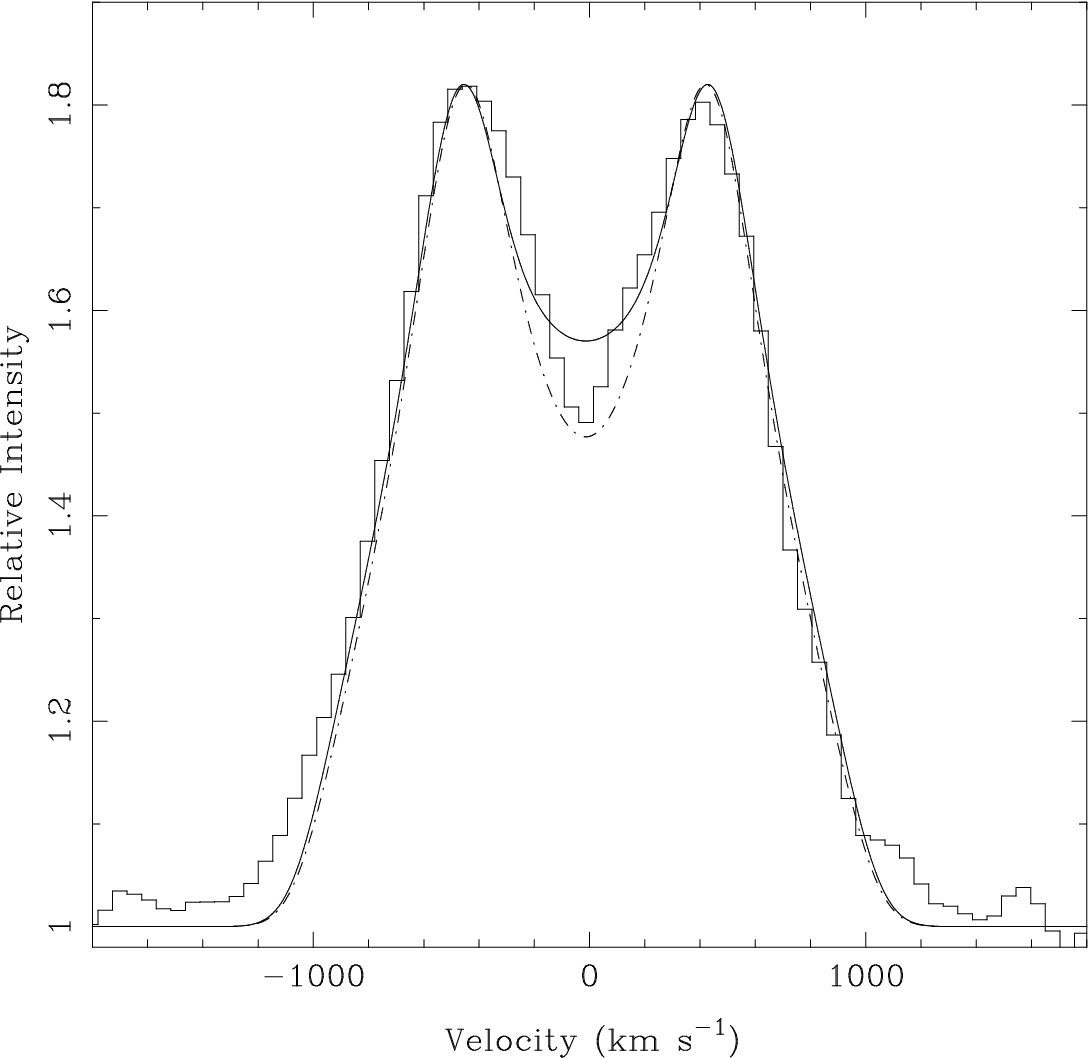}
\figcaption[f06.eps]{The mean H$\alpha$ profile is plotted as
the ``histogram''.  The $y$-axis scale is arbitrary, with the continuum
level at $y=1$.  Two model profiles are shown are an optically thin
model with $\alpha=2$, $r_1=0.17$, and $v_d=420$ km s$^{-1}$ (solid line)
and an optically thick model with 
$\alpha=2$, $r_1=0.17$, $v_d=420$ km s$^{-1}$, and $i=50^{\circ}$ 
(dash-dotted line).
\label{halphafig}}
\end{figure}

The XTE J1550-564 spectra show a strong, double-peaked H$\alpha$ emission
line.  
Fig.\ \ref{halphafig} shows the mean H$\alpha$ line profile obtained
by averaging the eleven spectra with the highest signal-to-noise.  The
equivalent width of the line is about 24.7~\AA, the full width
at half maximum is about 1500 km s$^{-1}$, 
and the maximum velocity
in the line wings is about 1000 km s$^{-1}$.  
We do not detect any feature
at the wavelength of H$\beta$, although the signal-to-noise is rather poor
there.  

The double-peaked emission line is characteristic of
emission from an accretion disk \citep{sma81}.   
In this case there is $\lesssim 8\%$ stellar contribution
to the H$\alpha$ profile since late G and early K stars have
relatively weak H$\alpha$ absorption lines (EW $<2$~\AA, Fig.\ 
\ref{fig3}). 
We modelled the line profile
using a simple model outlined in Smak (1981; see also Johnston, Kulkarni,
\& Oke 1989, and Orosz et al.\ 1994).  
The three main parameters are (1) the power-law exponent $\alpha$, which
determines the density function of the emitting atoms, $f(r) \propto
r^{-\alpha}$ (where the coordinate $r$ is normalized to 1 at the outer
edge of the disk), (2) the ratio
$r_1$  of the inner disk radius to the
outer disk radius, and (3)  the radial velocity $v_{d}$ of the outer
edge of the disk.  Horne \& Marsh (1986) have modified the model to
include profiles from an optically thick disk; this modification
requires that the inclination be included as an additional parameter.
An optically thin model with $\alpha=2$, $r_1=0.17$, and $v_d=420$
km s$^{-1}$ provides a reasonably good match to the observed profile 
(we have made no attempt to optimize the model fits),
although there is some excess emission in the blue line
wing and the central depression in the observed profile is deeper than the
model (see Fig.\
\ref{halphafig}).  
The optically thick models have deeper central depressions, and the
depth of an optically thick model with $i=50^{\circ}$ 
(and $\alpha=2$, $r_1=0.17$, and $v_d=420$
km s$^{-1}$, Fig.\ \ref{halphafig}) matches the depth of the central depression
reasonably well.  One should note that the inclination used in the model
profile may not be a good indicator of the actual binary inclination
(see \S \ref{inclcon}).
For example, the depth of the central depression in the H$\alpha$ profile
of GS 1124-683 (XN Mus91)
was observed to vary over an orbital cycle
\citep{oro94}.

The line wings of the H$\alpha$ profile of XTE J1550-564 are relatively
steep and do not extend to very high velocities (slightly more than
1000 km s$^{-1}$).  In contrast, the line wings of the 
H$\alpha$ profile of GS
1124-683 and H1705-250 (XN Oph77) extend out to almost 2000 km s$^{-1}$
\citep{oro94,rem96}.  
For A0620-00, the width of the H$\alpha$ profile has been
observed to vary between these extremes during 1991-1993 
\citep{oro94}.
The weakness or absence of the H$\beta$ line in XTE J1550-564 is 
similar to what is observed in H1705-250 in 1994 \citep{rem96}.  
There is some excess emission in the blue line wing of the XTE J1550-564
H$\alpha$ profile.  Excess emission in the blue line wing was observed
in GS 1124-683 in 1992, but not in 1993, and in A0620-00 in 1991, but
not in 1993.
Finally, we note
that the ratio of the disk velocity $v_d$ to the $K$-velocity in
XTE J1550-564 is $v_d/K_2=1.2$.  This value is well within the range
observed for the other black hole X-ray novae with (cool) low mass
companions (see Table \ref{vdisk}).  Thus in terms of the properties of
its quiescent
accretion disk, XTE J1550-564 appears to be a typical black
hole system.

\subsection{Ellipsoidal Light Curve Models and Derived 
Astrophysical Parameters}\label{inclcon}

The value of the orbital inclination $i$ is needed to compute the component
masses and other interesting parameters.
Normally one  models the ellipsoidal light curves
to estimate the inclination of a semi-detached system such as XTE J1550-564,
and for this purpose we
used the ELC code \citep{oro00}.
The model has many free parameters.  
The binary parameters for a circular orbit
are the inclination $i$, the mass ratio $Q$,
the orbital separation $a$, the orbital period $P$, and the
epoch of the inferior conjunction of the secondary star $T_0$(photo).
The
main parameters for the secondary star are its mean temperature
$T_{\rm eff}$, its gravity darkening exponent $\beta$, its Roche lobe
filling factor $f$, and its rotational velocity with respect to synchronous
$\Omega$.   Finally there are parameters related to the accretion disk
which are the outer radius of the disk $r_{\rm out}$, the inner radius of the
disk $r_{\rm in}$, the opening angle of the outer disk rim $\beta_{\rm rim}$,
the temperature at the inner edge $T_{\rm disk}$, and the power-law
exponent on the disk temperature profile $\xi$, where $T(r)=T_{\rm disk}
(r/r_{\rm in})^{\xi}$.  Some of these parameters can be
fixed at reasonable values.  We assume the secondary star exactly fills
its Roche lobe and is in synchronous rotation.  Hence $f=\Omega=1$.
Following \citet{cla00} the value of the
gravity darkening exponent $\beta$ is a function of the mean
temperature of the secondary star.  In this case
the gravity darkening exponent
is in the range of $0.06 \le \beta \le 0.11$ for
our adopted temperature range of
$4100 \le
T_{\rm eff}\le 5100$~K.
We used the specific intensities from the
{\sc NextGen} models \citep{hau99a,hau99b}, so consequently no parameterized
limb darkening law was needed.  The inner radius of the disk was taken
to be 3\% of the compact object's Roche lobe radius, which is roughly
$10^4$ Schwartzshild radii (see Hameury et al.\ 1997).

We are left with ten free parameters in the ellipsoidal model:
$Q$, $i$, $a$, $P$,
$T_0$(photo), $T_{\rm eff}$,
$\xi$, $T_{\rm disk}$, $r_{\rm out}$, and $\beta_{\rm rim}$.
For a given ellipsoidal model, we can compute various observable
quantities such as the shape of the light curves in a given bandpass,
the amplitude and phasing of the radial velocity curve of the
secondary star, the amount of disk light in various bandpasses,
the radius, gravity, and rotational velocity of the secondary star, etc.
The goal is to find the set of model parameters 
which
gives the best fit to all the observed properties of XTE J1550-564.

To accomplish the 
optimization
goal we used a genetic optimization code based on the PIKAIA
routine given in Charbonneau (1995), with the ``black sheep'' 
modification outlined by Bobinger (2000).
This genetic  code is very efficient at finding the global
optimal solution
in a large parameter space (see Charbonneau 1995 and Metcalfe 2001
for more detailed discussions of genetic algorithms).   
The algorithm used
in the code is quite simple.  We define a ``population'' of
100 random parameter sets.  For each parameter set we compute the
ellipsoidal model and define a ``fitness'' based on how well
the particular model compares with the available observations. 
After the fitness of the initial population is determined, 100 new 
parameter sets are produced as a result of ``breeding'' between pairs
of members, where the probability of breeding is based on the fitness.
Random variations (i.e.\ ``mutations'') are introduced into a small fraction
of the breeding events.  The process of breeding a new population and
evaluating its members is repeated (i.e.\ the population ``evolves'')
over many generations
(typically a few hundred or more) until convergence is achieved.

We will use the $\chi^2$ statistic to evaluate the goodness-of-fit
between the predicted and observed light and velocity curves:
\begin{eqnarray}
\chi^2_{\rm total} &=&  \chi^2_{\rm light} + \chi^2_{\rm velocity} \nonumber\cr
                   &=&  \sum_{j=1}^6\sum_{i=1}^{N_j}
             {(y(x_i;a_1\dots a_{10})-y_i)^2\over \sigma_i^2}
+ \sum_{i=1}^{17}
             {(y(x_i;a_1\dots a_{10})-y_i)^2\over \sigma_i^2}.
\end{eqnarray}
Here the notation $y(i_i;a_1\dots a_{10})$ means the ``model value
computed at $x_i$'' and $y_i$ means the ``observed quantity at the same
$x_i$.''
There is a double summation in the $\chi^2_{\rm light}$ term because
we have six filters.  
In the genetic code, the fitness of a population
member is related to its relative value of the $\chi^2$.  The
member with the lowest $\chi^2$ has the highest fitness, the one with
the second lowest $\chi^2$ has the second highest fitness, and so on.
We can also consider at least
three additional measures of fitness:  (i)~For each model
we can predict $\Theta$,
the duration of an X-ray eclipse, if any.  In the case
of XTE J1550-564, the X-ray eclipse duration is $\Theta=0$ at a high level of
confidence since
there were  no eclipses observed in the
almost 300 pointed {\em RXTE}
observations of
XTE J1550-564 during its  various outbursts nor in the extensive 
{\em RXTE} ASM observations
(Fig.\ \ref{fig0}b).  Thus we have:
\begin{equation}
\chi^2_{\Theta} =\cases{0, & if $\Theta=0$; \cr
                        10^6,& if $\Theta>0$. \cr}
\end{equation} 
As a result of this definition of $\chi^2_{\Theta}$, any model 
where the X-ray source is eclipsed is assigned a very low fitness.
(ii)~For each model we can predict $k_V$,
which is the fraction of the $V$-band light
contributed by the accretion disk.
Observationally, $k_V$ can be
estimated using the decompositions
discussed in \S\ref{secsec}.  
At 5500~\AA, the disk fraction is $\approx 0.4$ for the K3III template
(Fig.\ \ref{fig3}a), 
and $\approx 0.2$ for the G8 IV template
(Fig.\ \ref{fig3}b).
For the purposes of the ellipsoidal modelling we will adopt
$k_V=0.3\pm 0.1$.  We then have
\begin{equation}
\chi^2_{k_V}={(k_V(a_1\dots a_{10})-0.3)^2\over 0.1^2}.
\end{equation} 
For
simplicity we 
will neglect the small correlation between the disk fraction and the
effective temperature.  (iii)~For each model we can predict 
$V_{\rm rot}\sin i$, which is
the 
value of the projected rotational velocity of the secondary star.
In \S\ref{broadsubsec} we tentatively measured
$V_{\rm rot}\sin i=90\pm 10$ km s$^{-1}$.  We then have
\begin{equation}
\chi^2_{\rm rot}={(V_{\rm rot}\sin i(a_1\dots a_{10})-90)^2\over
10^2}.
\end{equation}
Our total $\chi^2$ becomes
$\chi^2_{\rm total}=\chi^2_{\rm light}
+\chi^2_{\rm vel}+\chi^2_{\Theta}+
\chi^2_{k_V} +\chi^2_{\rm rot}$.  

In any optimization procedure, one must decide how to assign weights
to the observations.  In our case
some initial fits were performed to get an approximate solution.
Based on these initial results,
the error bars on the photometric measurements were
scaled slightly to give $\chi^2_{\rm
light}/\nu\approx 1$ at the minimum in 
{\em each}
of the six filters
separately.  
Similarly, the errors on the radial velocities were
scaled to give $\chi^2_{\rm vel}/\nu\approx 1$ at the minimum.

Since our measured value of the rotational velocity may have
a large systemic error (the value we measured is smaller than
the spectral resolution), we ran two sets of fits.  Fitting run A
did {\em not} include the contribution of $V_{\rm rot}\sin i$ in 
the total $\chi^2$.
Fitting run B {\em did}   
include
the observed value of
$V_{\rm rot}\sin i=90$ km s$^{-1}$ in the total $\chi^2$
(i.e.\ we are taking the value of 90 km s$^{-1}$ at face value). 
For each of the two fitting runs A and B,
the genetic fitting code was run six
separate times on the data with the scaled
error bars, using a population size of 100.  
The volume of parameter space that was searched
was the same in all six cases, but
the order in which the parameters were encoded into the ``genes'' was
different.  In practice this means that the initial random
populations were different, and the way two population members
exchange parameters when breeding was different.
In all cases the final best-fitting parameters were
statistically identical after about 200 generations, 
giving us great confidence that our solutions
for each of runs A and B are truly the global ones.

\begin{figure}
\epsscale{0.7}
\plotone{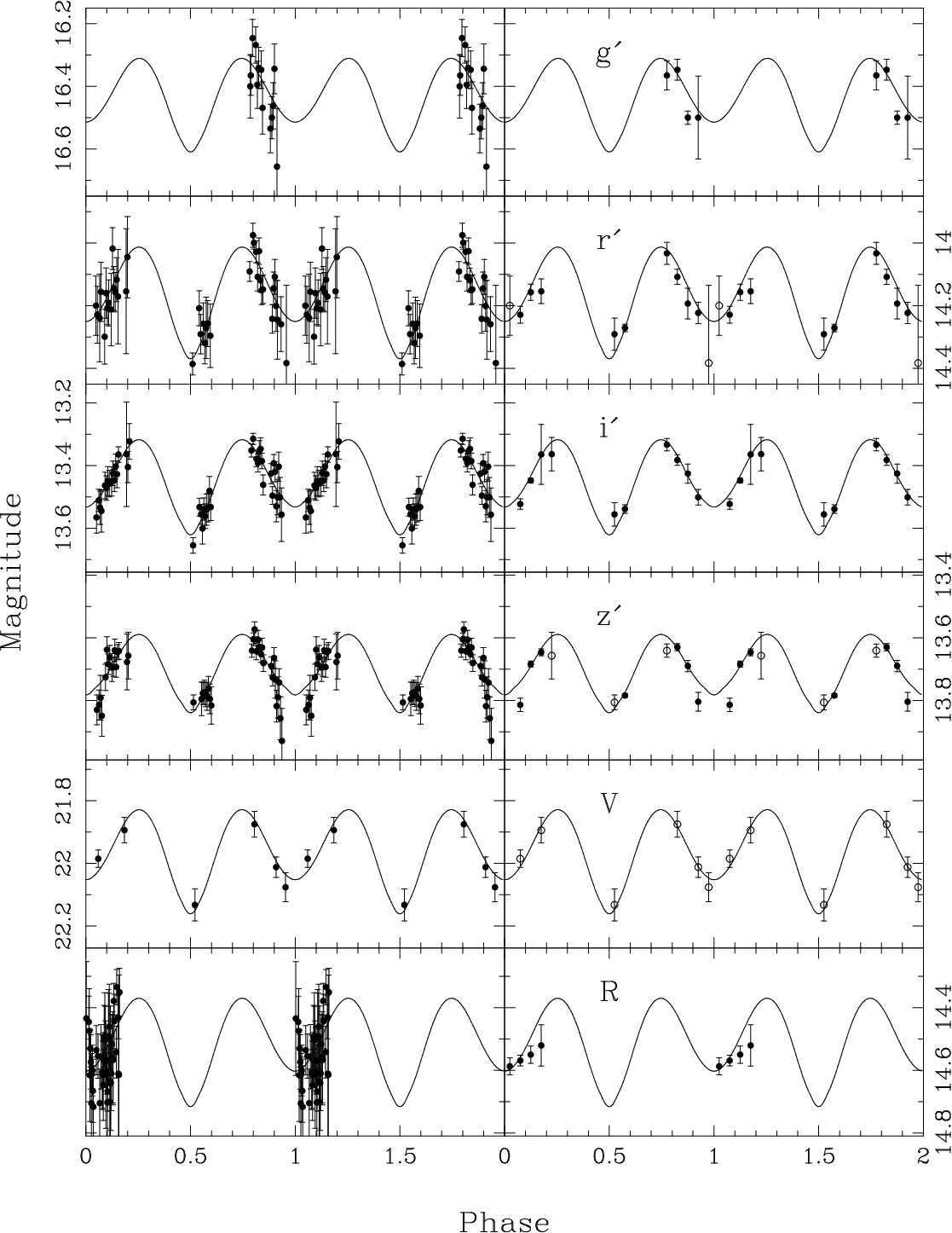}
\figcaption[f07.eps]{The left panels show the folded light
curves in the Sloan $g^{\prime}$,   $r^{\prime}$,   $i^{\prime}$,   and
$z^{\prime}$ filters and the Bessell $V$ and $R$ filters and
the best-fitting ellipsoidal models.  Note that only the $V$-band data
are calibrated onto the standard scale.
The smoothed light curves shown in the right panels were made by
computing the median magnitude
within bins 0.05 phase units wide.  Bins with
a single point are denoted by open circles.
\label{showlc}}
\end{figure}

The folded light curves in the six 
bands and
the model light curves are shown in Fig.\ \ref{showlc}.   
Although we fit the unbinned data, we also show in Fig.\ \ref{showlc}
smoothed versions of the light curves which were made by computing the 
magnitude within bins 0.05 phase units wide.  The smoothed light curves
generally follow the fitted models quite well.

In order to estimate the uncertainties on the fitted parameters and
the uncertainties on the derived astrophysical parameters
we collapse the 10-dimensional $\chi^2$ function onto the appropriate
parameter of interest.  
The 1, 2, and $3\sigma$ confidence limits are then the values of 
the parameter where $\chi^2=\chi^2_{\rm min}+1, \chi^2_{\rm min}+4$,
and  
$\chi^2_{\rm min}+9$, respectively.   
Using the genetic fitting code and the Black Sheep modification it
is quite easy to 
compute these confidence limits since the 
region of parameter space near the $\chi^2$ minimum is sampled quite well.
For each model that is computed, the $\chi^2$ of the fit, the
values of the free parameters, and the
computed astrophysical parameters (e.g.\ the black hole mass, the secondary
star radius, etc.)
are saved.  After a sufficiently large
number of models have been computed (more than 515,000 in fitting run A and
465,000 in fitting run B), 
we can collapse the $\chi^2$ function onto the parameter of
interest by 
simply plotting the 
$\chi^2$ vs.\ the parameter of interest and looking at the ``lower envelope''
(i.e.\ the minimum $\chi^2$ within small bins over the whole range).

\begin{figure}
\epsscale{0.7}
\plotone{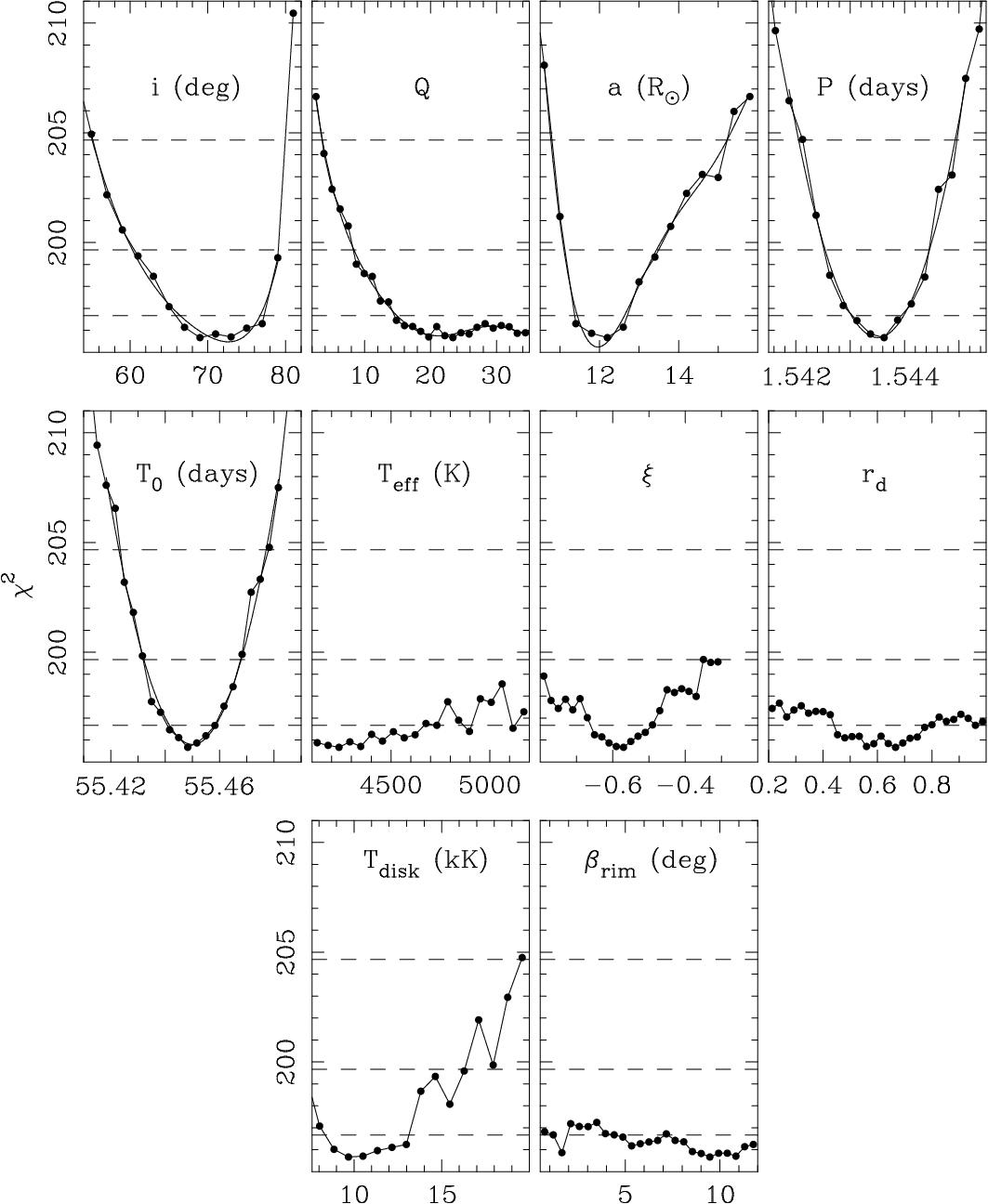}
\figcaption[f08.eps]{$\chi^2$ values as a function of the ten
fitting parameters for the fitting run A
(the run without a $V_{\rm rot}\sin i$
constraint).
They are, from left to right and top to bottom,
the inclination $i$ in degrees, the mass ratio $Q$, the orbital
separation $a$ in solar radii, the orbital period $P$ in days, the
phase zero-point $T_0$ in days since HJD 2~452~000, the mean temperature
of the secondary star $T_{\rm eff}$,
the power-law
exponent $\xi$ on the disk temperature profile, the radius of the
outer edge of the disk $r_d$
as a fraction of the compact object's Roche lobe
radius, the temperature of the inner edge of the disk
$T_{\rm disk}$ in units of 1000~K, and the opening angle of the outer
rim of the disk $\beta_{\rm rim}$ in degrees.  Fifth or sixth
order polynomial
fits to the first five curves  used to get the confidence regions
are shown as the smooth lines.
The dashed lines denote the 1, 2, and $3\sigma$ confidence limits.
\label{norotplotfitted}}
\end{figure}

\begin{figure}
\epsscale{0.7}
\plotone{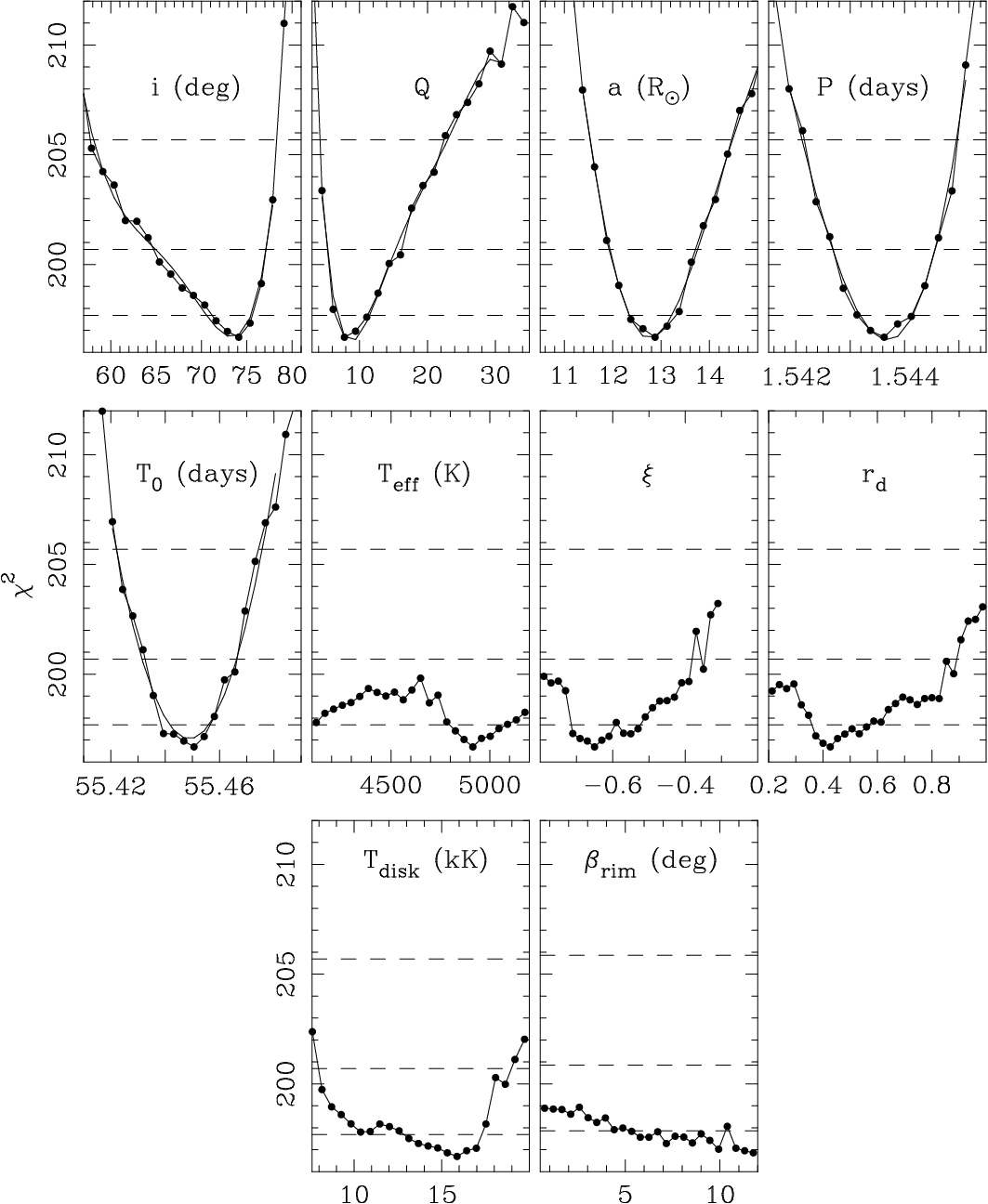}
\figcaption[f09.eps]{Similar to Fig.\ 
\protect\ref{norotplotfitted}, but for the fitting run B (the one with
$V_{\rm rot}\sin i=90\pm 10$ km s$^{-1}$ as a fitting constraint).
\label{plotfitted}}
\end{figure}

Fig.\ \ref{norotplotfitted}
shows the collapsed $\chi^2$ function for the ten fitted parameters
in the fitting run A, and Fig.\ \ref{plotfitted}
shows the similar plot for the fitting run B.
In some cases, we
fitted
fifth or sixth
order polynomials to the curves to more objectively
to determine the value of the parameter
where $\chi^2=\chi^2_{\rm min}$ and the 1, 2, and $3\sigma$ confidence
limits.
In general, the important geometrical parameters in the ellipsoidal model
(i.e.\ the inclination $i$, separation $a$, etc.) have well-defined
minima and
$3\sigma$ limits.  In addition, the derived parameter ranges 
are not too different
between fitting run A and fitting run B.
The one exception is the mass ratio $Q$.  In the
fitting run A, where the observed rotational velocity was not included
in the total $\chi^2$, the value of $Q$ can only be assigned a lower
limit ($Q>6$ at the $3\sigma$ level).  However, the mass ratio $Q$ does
have well defined upper and lower limits in the fitting run B, where
the observed rotational velocity was included in the total $\chi^2$.
On the other hand, the accretion disk parameters and the effective
temperature of the secondary star are not
so well constrained.  

\begin{figure}
\epsscale{0.7}
\plotone{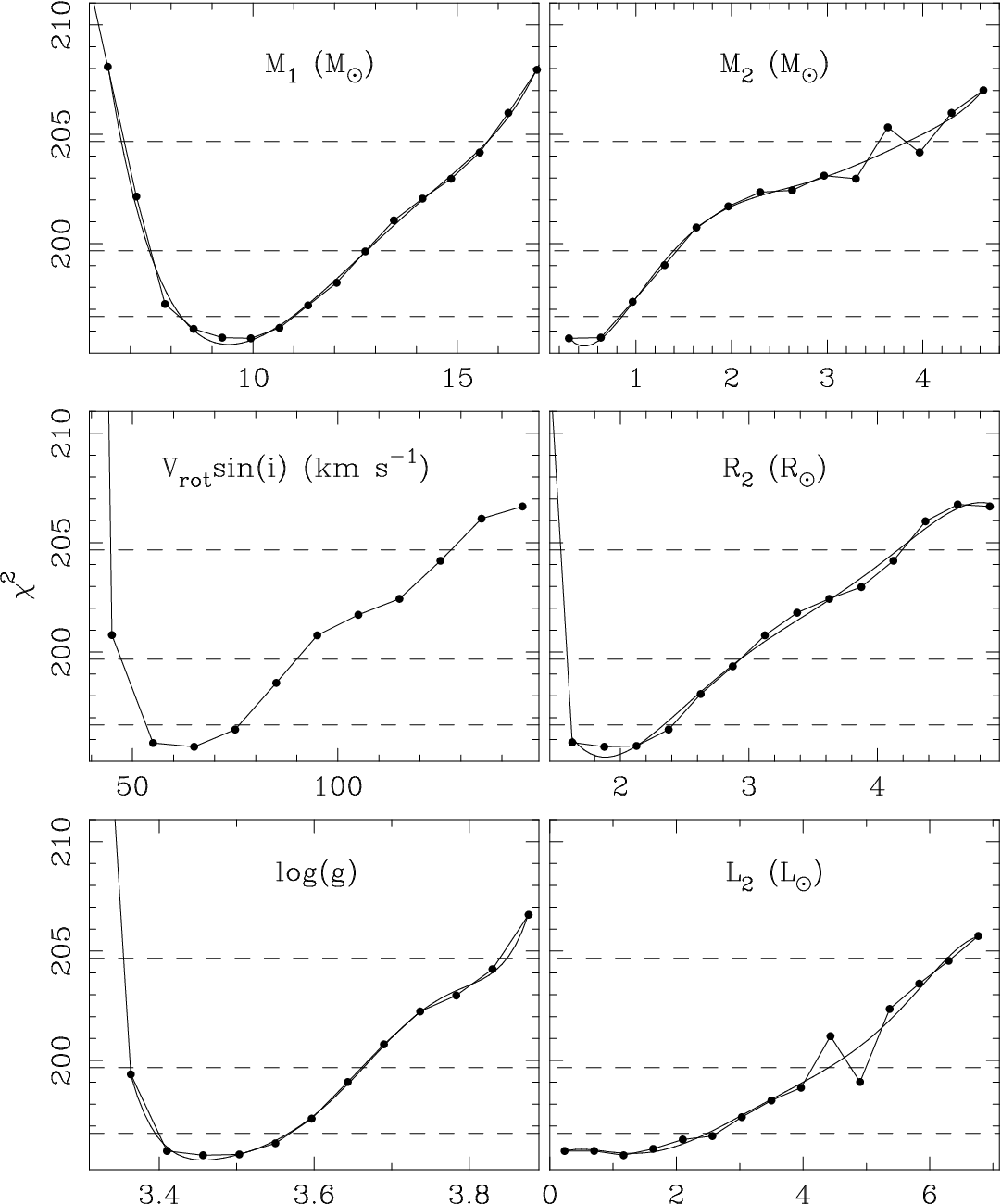}
\figcaption[f10.eps]{$\chi^2$ values as a function of various
derived parameters for fitting run A
(the run without a $V_{\rm rot}\sin i$
constraint).
They are the black hole mass $M_1$ in solar masses, the secondary
star mass $M_2$ in solar masses, the predicted projected rotational velocity 
of the secondary star $V_{\rm rot}\sin i$ in km s$^{-1}$, the radius
of the secondary star $R_2$ in solar radii, the surface gravity
of the secondary star $\log g$ in cgs units, and the bolometric luminosity
of the secondary star $L_2$ in solar units.  Fifth 
or sixth order polynomial
fits
to the curves  used to get the confidence regions
are shown as the smooth lines.
The dashed lines denote the 1, 2, and $3\sigma$ confidence limits.
\label{norotplotderiv}}
\end{figure}

\begin{figure}
\epsscale{0.7}
\plotone{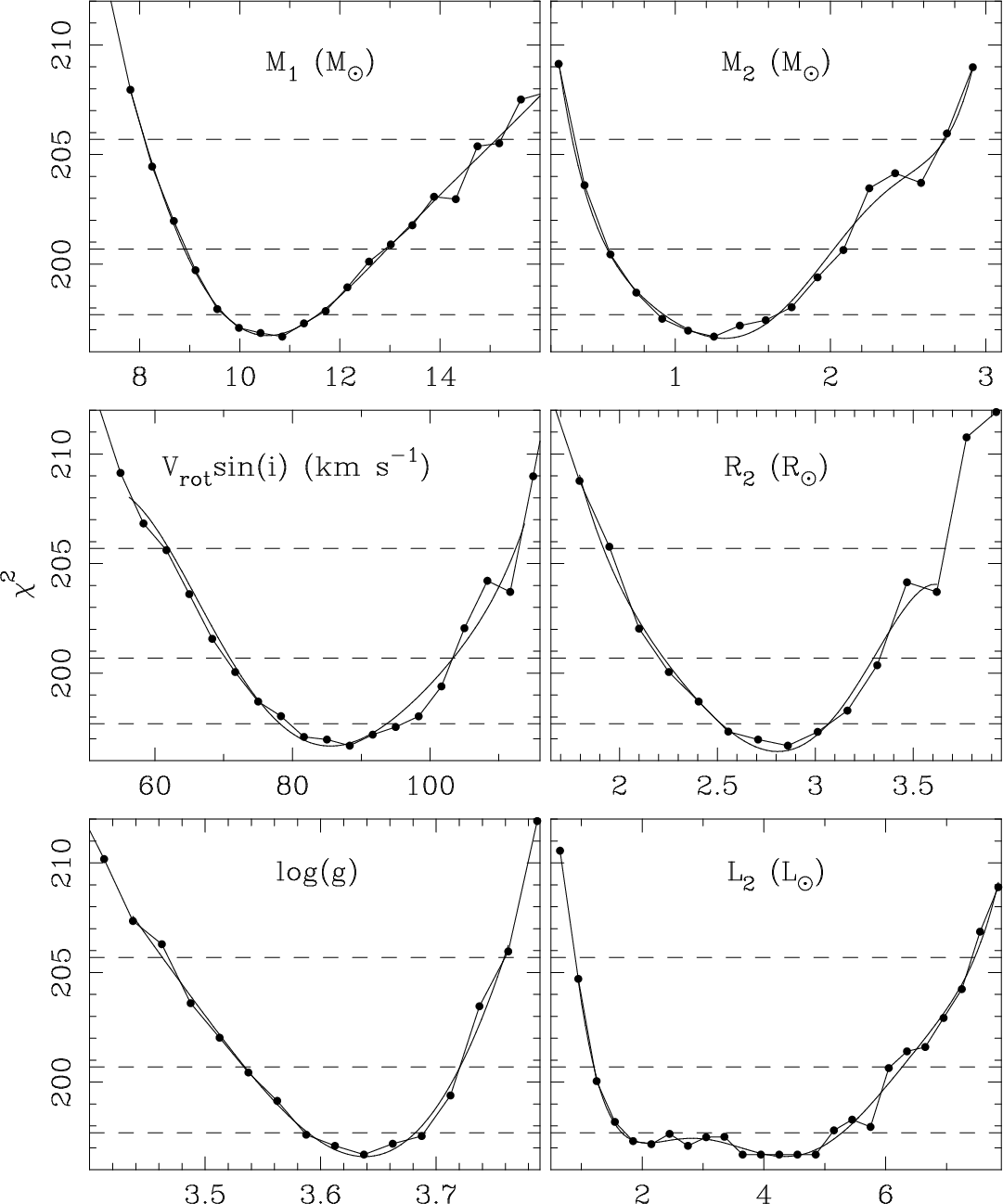}
\figcaption[f11.eps]{Similar to Fig.\ 
\protect\ref{norotplotderiv}, but for fitting run B (the one with
$V_{\rm rot}\sin i=90\pm 10$ km s$^{-1}$ as a fitting constraint).
\label{plotderiv}}
\end{figure}

Fig.\ \ref{norotplotderiv} shows the collapsed $\chi^2$ function 
for six derived 
parameters in fitting run A:  
$M_1$ (the compact object's mass), and $M_2$ (the 
secondary star's mass),
$V_{\rm rot}\sin i$,
(the predicted rotational velocity of the secondary star), 
$R_2$ (secondary star's radius), $\log g$ (the secondary star surface
gravity), and $L_2$ (the secondary star's bolometric luminosity).  
Fig.\ \ref{plotderiv} shows the similar plot for fitting run B.
Key results are summarized are in Table \ref{parm}.
The nominal mass of the black hole is
$9.41\,M_{\odot}$ for fitting run A and $10.56\,M_{\odot}$
for fitting run B, respectively.    We will adopt
the results from fitting run B in the following discussions.

The only observational constraint related to the disk
at present is that the disk should contribute roughly 30\% of the light
in the $V$-band, and there is a wide range of accretion disk parameters
which produce a disk with roughly the correct brightness.  Since the 
inclination seems to be fairly high, the disk may partially eclipse the
secondary star, and vice versa.  
Unfortunately, we do not have complete phase coverage, so we cannot
at present detect the subtle signatures of grazing eclipses.  
Therefore, observations with much better phase coverage near the 
conjunction phases should be obtained.  Alternatively, 
observations in the infrared (i.e.\ $J-$,
$H-$, and $K-$ bands) would also 
be especially useful since the usual assumption
is that the relative contribution of the accretion disk is much smaller
at these wavelengths.  
This assumption can and should be tested with
good quality infrared spectra.  If the disk turns out to be faint in the
infrared, then the model fits would of course be much less sensitive to the
assumed disk parameters.

The inclusion of the
rotational velocity as an additional fitting constraint
resulted in only a modest ($\approx 10\%$) change in
the derived mass of the black hole.  This is because the
black hole mass scales as $f(M)(1+1/Q)^2$, and the mass
ratio $Q$ is relatively large.  
On the other hand, the derived properties
of the secondary star are
quite different between the two fitting runs.  For example, 
$M_2< 0.79\,M_{\odot}$ ($1\sigma$) for fitting run A and
$M_2=1.31\,M_{\odot}$ for fitting run B.
Clearly,
a better measurement
of $V_{\rm rot}\sin i$ would be quite useful since it would place a much
stronger
constraint on the mass ratio, which in turn would place stronger
constraints on the properties of the secondary star.  In order 
to significantly
improve the precision of the black hole mass, one would need to 
also improve the precision of the mass function $f(M)$.

Finally, it is worthwhile to point out that the
results summarized in
Table \ref{parm} are our best estimates based on the data we
currently have.  Although the parameter uncertainties have been
rigorously computed, they represent {\em statistical} errors only.
There may be systematic errors in, for example, the measured
rotational velocity $V_{\rm rot}\sin i$ owing to the low spectral resolution
or in the derived inclination owing to the less than perfect phase
coverage and errors in measuring the contamination from the accretion disk.
It was straightforward to perform two fitting sequences to see what
effect including the rotational velocity as a fitting constraint had
on the outcome.  It is less straightforward to quantify the effects
of incomplete phase coverage or imperfect disk/stellar decomposition.
However, we can say with some confidence that since our derived inclination 
is already quite high and near the upper limit
imposed by the lack of X-ray eclipses, 
it is unlikely that we have underestimated
the true inclination significantly.  
Ultimately one would need to obtain additional data to see if
similar light curves are observed.

\subsection{Distance}\label{dsubsec}

The distance to the source is  a function of four parameters:
The mass $M_2$ and temperature $T_{\rm eff}$ of the
secondary star fix its luminosity (see Eq.\ \ref{den}); interstellar
extinction ($A_V$) makes the source appear fainter; and extra light
from the accretion disk (parameterized by the $V$-band ``disk
fraction'' $k_V$) makes the source appear brighter.  
We have reasonably good values for
all  of these parameters.  
As noted in \S\ref{secsec},
$A_V \approx 4.75$ 
can be derived by modelling the X-ray spectrum
and assuming a typical gas to dust ratio for the interstellar
medium; $k_V
\approx 0.30$ was determined from our spectral decomposition (Figs.\
\ref{fig3}a and \ref{fig3}b);
and a temperature range of
$4100\le T_{\rm eff} \le 5100$ K was deduced for
the secondary star.
Using these results
and a simple Monte Carlo procedure, we computed the distance and its
uncertainty for various values of the secondary star mass $M_2$.  
We 
used the synthetic photometry computed 
from the {\sc NextGen} 
models\footnote{ftp://calvin.physast.uga.edu/pub/NextGen/Colors/} 
to determine the expected
absolute $V$ magnitude of the star from its
temperature, radius, and surface gravity.
For
this procedure we adopted $4100\le T_{\rm eff}\le 5100$~K, $k_V = 0.30\pm
0.05$, a mean apparent $V$ magnitude of $22.0\pm 0.2$, and $N_H=(8.5\pm
2.3)\times 10^{21}$ cm$^{-2}$ (90\% confidence) and
$A_V=N_H/1.79\times 10^{21}$ (exact).  The results are displayed in
Table \ref{tabdist} for a wide range of secondary star masses.  
For masses in the range of 0.15 to
$3.0\,M_{\odot}$ the computed distance is in the range 3.0 to 7.6
kpc.  However, at a fixed mass, the formal error on the distance is
rather large due to the relatively large adopted temperature range and
to the relatively large error in $N_H$.  
The distance for the case 
of $M_2=1.31\,M_{\odot}$ is 5.9 kpc, again with a rather large
formal error ($1\sigma$ range of 2.8 to 7.6 kpc).

We can make an independent estimate of the distance
if we assume the 
observed systemic velocity of $\gamma=-68\pm 12$
km s$^{-1}$ is entirely due to differential galactic rotation.
We use the rotation curve of Fich, Blitz, \& Stark (1989) and the
standard IAU rotation constants of
$R_0=8.5$ kpc and $\Theta_0=220$ km s$^{-1}$.
The rotation curve for a source with
galactic coordinates 
of $\ell-325.88^{\circ}$ and $b=-1.83^{\circ}$ is roughly
parabolic, with a minimum radial velocity of
$\approx -95$ km s$^{-1}$ at a distance of $\approx 7$
kpc.  The expected distance for a source with
$\gamma=-68$ km s$^{-1}$ is either $\approx 4.4$ kpc
or $\approx 9.7$ kpc.  Considering the errors on
$\gamma$, the kinematic distance can be anywhere
between $\approx 3.2$ kpc and $\approx 10.8$ kpc
at the $1\sigma$ level, which is entirely consistent with the
ranges derived above.  Given the general agreement between
the two distance estimates we conclude that XTE J1550-564 has
essentially no (radial) velocity with respect to its local standard
of rest.

We can use the distance estimates for XTE
J1550-564 to calculate an order of magnitude estimate for the maximum
X-ray luminosity seen during outbursts.  The spectral analysis of
\citet{sob00}
for the intense flare of 1998 September 19
implies an unabsorbed flux of $3.7 \times 10^{-7}$ erg cm$^{-2}$
s$^{-1}$ over the range of 1-20 keV, with the spectrum dominated by
the power-law component. 
For our favored distance of 5.3 kpc, the
isotropic luminosity is then 
$1.77\times10^{39}$ erg s$^{-1}$
(with a large error owing to the uncertainty in the distance).  
This is just slightly above
the Eddington luminosity for a $10.56\,M_{\odot}$ black hole, which is a
remarkable coincidence given the significant uncertainty in the
distance.

\section{Discussion}\label{dissec}

\citet{bai98} analyzed the distribution of black hole masses in seven
systems and found a high probability that six of the seven systems
(GRO J0422+32, A0620-00, GS 1124-683, GRO J1655-40, H1705-250,
and GS 2000+25) have masses which are consistent
with  
$\approx 7\,M_{\odot}$.  The seventh system, V404 Cyg, has a mass 
which is significantly
larger (about $12\,M_{\odot}$; Shahbaz et al.\ 1994).   
The mass of
the black hole in XTE J1550-564 is well above $7\,M_{\odot}$. With
$M_1\ge 8.1\,M_{\odot}$ at $3\sigma$ confidence (Table \ref{parm}), its mass
may
in fact be similar to the mass of the black hole in V404 Cyg.   
Since the work of \citet{bai98}, the mass functions for
seven additional systems have been measured, and improved parameters
for some of the original seven systems have been measured.  Thus the issue
of the observed black hole mass distribution should be revisited to see
if the clustering of black hole masses near $7\,M_{\odot}$ 
is still significant.
Given the new data,
we can begin to make
meaningful comparisons with formation theory.  For example,
the detailed formation models of \citet{fry01} predict a mass distribution
which is continuous and which extends over a broad range (in particular,
they predict no peak at $\approx 7\,M_{\odot}$).

The determination of the black hole mass
for XTE J1550-564 is especially important for the interpretation of
the high-frequency X-ray quasi-periodic oscillations (QPOs) observed
for this system (Remillard et al. 2002; Homan et al. 2001).  Models
for several types of oscillations predicted in general relativity are
under investigation as possible causes of these QPOs (e.g. Remillard
2001, and references therein); all of these depend on both the mass and
spin of the black hole, and possibly also on conditions in the inner
accretion disk.
Despite considerable uncertainties in the models, we can
offer a few comments on the implications of our mass determination for
XTE J1550-564.  At the nominal mass of  
$10.7\,M_{\odot}$,
oscillations at the frequency of the last stable orbit for a
Schwarzschild black hole (spin parameter, $a_* = 0$) would be seen at a
frequency, $\nu = 2199 / (M_{\rm BH} / M_{\odot}) = 208.2 $ Hz \citep{sha83}.
Since XTE J1550-564 has exhibited QPOs
with frequencies up to 284 Hz \citep{hom01}, it appears that
plausible mechanisms require $a_* > 0$.
(The oscillation frequency would be 284 Hz (for $a_*=0$) if
$M_{\rm BH}=7.74\,M_{\odot}$, which is below the $3\sigma$ lower limit
on the black hole mass.)
This argument was used by 
\citet{str01}
to suggest that the
450 Hz QPO in GRO J1655-40 implies appreciable spin for the black hole
in that system. 
We further note that the ratio of the black hole masses for XTE
J1550-564 ($M_1=10.6\,M_{\odot}$, Table \ref{parm})
and GRO J1655-40 ($M_1=6.3\,M_{\odot}$, Greene, Bailyn, \& Orosz 2001)
is $\sim 1.7$, which is nearly
inversely
proportional to their maximum QPO frequencies: 
$(284 {\rm Hz}/{\rm 450 Hz})^{-1}\sim 1.6$.  
We may then
speculate that these results are consistent with a common QPO origin,
with QPO frequencies that vary as $M^{-1}$, which might be expected
if the black holes have similar values of the spin parameter.

Both XTE J1550-564 and GRO J1655-40 have had several closely-spaced outburst
events relatively shortly after their initial discoveries.  Interestingly
enough, optical precursors to X-ray outbursts (i.e.\ X-ray delays)
have been observed
for both sources.  
According to \citet{ham97},
these X-ray delays can best be understood using a two component accretion
model
consisting of a standard thin disk in the outer regions and an
ADAF region in the inner region.  At the onset of an outburst, the heating
front reaches the ADAF region relatively quickly, but then must propagate 
more slowly on a viscous timescale, thereby greatly delaying the
production of X-rays.    
\citet{oro97a} observed an optical precursor to the
1996 April outburst of GRO J1655-40.  In that case, the delay
between the $V$-band rise and the detection by the ASM
on {\em RXTE} was $5.6\pm 0.8$ days.  In the case of the 2000
outburst of XTE J1550-564,
\citet{jai01c} reported a delay of $8.8\pm 0.6$ days between the
rise in $V$ and the detection by the {\em RXTE} ASM.  The ratio
of these delays is $8.8/5.6\sim 1.6$, which is nearly the same as
the ratio of the masses and is the same as the 
inverse ratio of their maximum
QPO frequencies ($\sim 1.7$ and $\sim 1.6$ 
respectively).    We may again speculate
that these results imply that the size of the ADAF cavity scales
with the mass of the black hole.

For the case of a companion star
mass of $1.3\,M_{\odot}$ (Table \ref{parm}),
the companion star in XTE J1550-564 appears to have 
mass that is  larger than the typical mass
for
X-ray  novae with a cool (K or M) companion.
For comparison, the K0 secondary star in V404 Cyg has a mass of
about $0.7\,M_{\odot}$ \citep{sha94}.  Since the secondary star
in XTE J1550-564 is about three times larger than a normal K3/4 dwarf, 
its evolutionary state may be different than the dwarf secondaries 
in the short
period X-ray novae such as A0620-00 or GS 1124-683.  If the star
is still in the core hydrogen burning phase, then its nuclear evolution
timescale would be $\approx 1\times 10^9$ years, leading to an average
mass transfer rate of $10^{-9}\,M_{\odot}$ yr$^{-1}$.   If the star
has finished its core hydrogen burning, then presumably its radius
is expanding on a thermal timescale of $\approx 10^{7}$ years, leading
to an average mass transfer rate of $10^{-7}\,M_{\odot}$ yr$^{-1}$.

We can make a rough estimate of the mass transfer rate in XTE
J1550-564 as follows.  The {\em RXTE} satellite performed numerous
observations of XTE J1550-564 during its 1998 and 1999
outbursts. \citet{sob00} give the unabsorbed 2-20 keV flux and the
20-100 keV flux for the $\approx 200$ PCA and HEXTE observations
from this period,
respectively (their Fig.\ 6).  Integrating under the curves, we find a
total 2-100 keV fluence of 1.06 erg cm$^{-2}$ for the 1998-1999
outburst.  Integrating the ASM light curve (Fig.\ 1) we find that the
outburst in 2000 was about 20 times fainter in the 2-12 keV band than
the 1998 and 1999 events.  Assuming similar spectral characteristics for
the two events, we conclude that total X-ray fluence in the 2-100 keV
band was 1.1 erg cm$^{-2}$ in the interval between 1998 and 2001.  To
get the average mass transfer rate, we need to know the recurrence
time between major outbursts and the distance.  According to Chen,
Shrader, \& Livio (1997), the sky coverage provided by scanning X-ray
instruments has been quite good since about 1988, so it seems likely
that the recurrence time of XTE J1550-564 is at least ten years.  The
distance is not well known, but it is probably about 6 kpc to within a
factor of two (Table \ref{tabdist}).  In Table \ref{tabenergy} we give
total isotropic energies and average mass transfer rates for assumed
distances of 3, 6, and 10 kpc and recurrence times of 10 and 50 years
(we assume an accretion efficiency of 10\% and that the mass transfer
rate from the secondary star is the same as the mass transfer rate
onto the black hole).  The mass transfer rates
are between about $1\times 10^{-10}$ and $7\times 10^{-9}\,M_{\odot}$
yr$^{-1}$, which suggests the star is evolving on a relatively long
nuclear timescale.

GRO J1655-40 might be 
a similar system in terms of its evolutionary state.
The average mass transfer rate in GRO J1655-40 has
been estimated to be about $1\times 10^{-10}\,M_{\odot}$ yr$^{-1}$
(van Paradijs 1996).
However, the secondary star  is located in the
Hertzsprung gap in the Hertzsprung-Russell
diagram (e.g.\ Orosz \& Bailyn 1997),  
so one would expect a mass transfer rate of about $1\times 10^{-7}\,M_{\odot}$
yr$^{-1}$ (e.g.\ Kolb et al.\ 1997).
Regos, Tout, \& Wickramasinghe (1998) have suggested that the
secondary star in GRO J1655-40
is still in the core hydrogen burning
phase, and they have constructed evolutionary models with mass
transfer rates much closer to the observed value.
It remains to be seen if a similar model can be constructed
for XTE J1550-564.

Although the presence of a black hole in XTE J1550-564 is now firmly
established, follow-up observations would be desirable to confirm and
improve upon our results.   The statistical error in the mass function can
be easily reduced by adding more radial velocities.  
Spectroscopic observations with higher resolution
should be obtained so that the rotational velocity of the companion star
can be measured more precisely.  Our inclination limits can  be
improved by obtaining multicolor light curves with much better
phase coverage.

\section{Summary}\label{sumsec}

Our
optical spectroscopic observations of the companion star in 
XTE J1550-564 have established the orbital period
of $P_{\rm sp}=1.552\pm 0.010$ days and the 
semiamplitude of $K_2=349\pm 12$ km s$^{-1}$.  The optical mass
function is $f(M)=6.86\pm 0.71\,M_{\odot}$. The rotational velocity of
the companion star is $V_{\rm rot}\sin i=90\pm 10$ km s$^{-1}$, which
implies a mass ratio of $Q\equiv M_1/M_1=6.6^{+2.5}_{-1.6}$ 
($1\sigma$).  This result should be confirmed with better data.
From light curve modelling we find 
the most likely value of
the mass of the compact object is $9.41\,M_{\odot}$ with a $1\sigma$
range of $8.36\,M_{\odot}\le M_1\le 10.76\,M_{\odot}$.
If we apply our tentative value of $V_{\rm rot}\sin i=90\pm 10$ km s$^{-1}$
as an additional constraint
in the ellipsoidal modelling, we find
the most likely value 
of
the mass of the compact object is $10.56\,M_{\odot}$ with a $1\sigma$
range of $9.68\,M_{\odot}\le M_1\le 11.58\,M_{\odot}$.
The black hole nature of XTE J1550-564 is therefore firmly established
by this work.

\acknowledgments

It is a great pleasure to
thank the numerous people at the Las Campanas, Paranal,
and La Silla  Observatories who
made this project a success, and the two YALO observers, Juan Espinoza and
David Gonzalez Huerta, for providing data in a timely and efficient manner. 
JAO appreciates useful discussions with
Norbert Langer.  CDB and RKJ were supported in part by the NSF grant
AST-9730774 and JEM was supported in part by NASA grant NAG5-10813.
PJG was supported by a CfA Fellowship.
RR acknowledges support from the NASA contract
to MIT for instruments of the {\em RXTE} mission.
This research has made use of 
the SIMBAD database, operated at CDS, Strasbourg, France, and NASA's 
Astrophysics Data System Abstract Service.

%% Generally speaking, only the figure captions, and not the figures
%% themselves, are included in electronic manuscript submissions.
%% Use \figcaption to format your figure captions. They should begin on a
%% new page.

\clearpage

%% No more than seven \figcaption commands are allowed per page,
%% so if you have more than seven captions, insert a \clearpage

%% There must be a \figcaption command for each legend. Key the text of the
%% legend and the optional \label in curly braces. If you wish, you may
%% include the name of the corresponding figure file in square brackets.
%% The label is for identification purposes only. It will not insert the
%% figures themselves into the document.
%% If you want to include your art in the paper, use \plotone.
%% Refer to the on-line documentation for details.

%% Tables should be submitted one per page, so put a \clearpage before

%% Two options are available to the author for producing tables:  the
%% deluxetable environment provided by the AASTeX package or the LaTeX
%% table environment.  Use of deluxetable is preferred.
%%

%% Three table samples follow, two marked up in the deluxetable environment,
%% one marked up as a LaTeX table.

%% In this first example, note that the \tabletypesize{}
%% command has been used to reduce the font size of the table.
%% Note also that the \label command needs to be placed 

\clearpage

\begin{deluxetable}{lr}
%\tabletypesize{\scriptsize}
\tablecaption{Orbital Parameters for XTE J1550-564. \label{tab2}}
\tablewidth{0pt}
\tablehead{
\colhead{Parameter} & \colhead{Value}
}
\startdata
$P_{\rm sp}$, spectroscopic orbital period (days) & $1.552\pm 0.010$ \cr
$P_{\rm ph}$, photometric orbital period (days)   & $1.5437\pm 0.0005$ \cr
$K_2$ velocity (km s$^{-1}$)         & $349\pm 12$    \cr
$\gamma$ velocity (km s$^{-1}$)         & $-68\pm 19$    \cr
$T_0$, spectroscopic\tablenotemark{a} ~(HJD 2,452,000+) & $54.296\pm 0.014$ \cr
$T_0$, spectroscopic\tablenotemark{b} ~(HJD 2,452,000+) & $55.448\pm 0.010$\cr
Mass function ($M_{\odot}$)  & $6.86\pm 0.71$ \cr
\enddata
\tablecomments{All quoted uncertainties are $1\sigma$.}
\tablenotetext{a}{The time of the maximum radial velocity of the 
secondary star.}
\tablenotetext{b}{The time of the inferior conjunction of the secondary star.}
\end{deluxetable}

%\clearpage

\begin{deluxetable}{ccc}
%\tabletypesize{\scriptsize}
\tablecaption{Disk Velocities for Black Hole Systems. \label{vdisk}}
\tablewidth{0pt}
\tablehead{
\colhead{Object Name} & \colhead{$v_d/K_2$}
& \colhead{Reference}
}
\startdata
H1705-250    & 1.11  & 1 \cr
GS 1124-683 & 1.13  & 2 \cr
GRO J0422+32 & 1.17 & 3,4 \cr
XTE J1550-564 & 1.20 & 5 \cr
A0620-00\tablenotemark{a}     & 1.24  & 2 \cr
GRS 1009-45  & 1.26  & 6 \cr
V404 Cyg     & 1.33 & 7,8 \cr
GS 2000+25   & 1.35  & 9 \cr
A0620-00\tablenotemark{b}     & 1.47  & 2 \cr
\enddata
\tablecomments{References: 
(1)~Harlaftis et al.\ 1997;
(2)~Orosz et al.\ 1994;
(3)~Orosz \& Bailyn 1995;
(4)~Filippenko, Matheson, \& Ho 1995;
(5)~This work;
(6)~Filippenko et al.\ 1999;
(7)~Casares, Charles, \& Naylor 1992;
(8)~Gotthelf et al.\ 1992;
(9)~Filippenko, Matheson, \& Barth 1995}
\tablenotetext{a}{Observed 1991 profile}
\tablenotetext{b}{Observed 1993 profile}
\end{deluxetable}

%\clearpage

\begin{deluxetable}{lcccccc}
\tablecaption{Fitted and Derived
Astrophysical Parameters for XTE J1550-564
from Light Curve Modelling\label{parm}}
\tablewidth{0pt}
\tablehead{
\colhead{} & \multicolumn{3}{c}{(No $V_{\rm rot}\sin i$ constraint)}
&
\multicolumn{3}{c}{($V_{\rm rot}\sin i=90\pm 10$ km s$^{-1}$)}
\\
\colhead{Parameter} & \colhead{Central} &
\colhead{$1\sigma$ range} & \colhead{$3\sigma$ range}
& \colhead{Central} &
\colhead{$1\sigma$ range} & \colhead{$3\sigma$ range} \\
\colhead{} & \colhead{Value} & \colhead{} & \colhead{} & \colhead{Value}
& \colhead{} & \colhead{}
}
\tablecolumns{7}
\startdata
Inclination (degrees)               & 72.6 & 67.0-77.4& 55.2-79.0  
                                               & 73.5 & 70.8-75.4 & 58.5-78.4          \cr
Mass ratio                          & $\approx 21.0$ & $>12.0$ & $>3.6$ 
                                               & 8.5 & 6.7-11.0 & 4.5-22.5             \cr
Orbital separation ($R_{\odot}$)    & 11.96 & 11.52-12.50 & 10.79-15.09
                                               & 12.77 & 12.35-13.22 & 11.53-14.45     \cr
Orbital Period (days)               & 1.5435 & 1.5430-1.5440 & 1.5421-1.5449 
                                               & 1.5437& 1.5432-1.5441 &1.5421-1.5450  \cr
$K_2$ (km s$^{-1}$)                 & 359.4  & 350.2-368.6 & 331.2-387.4 
                                               & 361.2 & 352.3-370.1 & 333.8-388.2     \cr 
                                    &        &             &  
                                               &       &             &                 \cr
Black hole mass ($M_{\odot}$)       & 9.41  & 8.36-10.76 & 6.81-15.60 
                                               & 10.56 & 9.68-11.58 & 8.10-15.15     \cr
Secondary star mass ($M_{\odot}$)   & $\approx 0.4$ & $<0.79$  & $<4.10$ 
                                               & 1.31 & 0.94-1.64  & 0.34-2.73     \cr
Total mass ($M_{\odot}$)            & 9.62 & 8.71-10.86 & 7.29-14.83 
                                               & 11.70 & 10.70-12.89 & 8.85-16.84     \cr
Secondary star radius ($R_{\odot}$) & $\approx 1.88$ & 1.62-2.24 & 1.63-4.06 
                                    & 2.81 & 2.55-3.04 & 1.93-3.61         \cr
Secondary star luminosity ($L_{\odot}$)  & $\approx 1.4$ & $<2.5$  &  $<6.3$
                                    & $\approx 4.0$ & 1.7-5.2  &  1.0-7.4 \cr
Secondary star gravity ($\log g$) & 3.46 & 3.40-3.56  &  3.36-3.84
                                    & 3.63 & 3.59-3.68  &  3.46-3.76 \cr
\enddata
\end{deluxetable}

%\clearpage

\begin{deluxetable}{ccccc}
\tablecaption{Distance as a Function of Mass\label{tabdist}}
\tablewidth{0pt}
\tablehead{
\colhead{Assumed Secondary} & \colhead{Radius}       & \colhead{$M_V$ for}  &
\colhead{Distance (kpc)}  &  \colhead{$1\sigma$ Distance Range (kpc)} \\
\colhead{Mass ($M_{\odot}$)} & \colhead{($R_{\odot}$)} & \colhead{$T_{\rm eff}
=4600$~K} &
\colhead{for $T_{\rm eff}
=4600$~K}     & 
\colhead{for $4100\le T_{\rm eff}
\le 5100$~K}   
}
\startdata
0.15   & 1.47 & 5.25  &  3.0  & 1.4--3.9 \cr
0.50   & 2.20 & 4.44  &  4.4  & 2.1--5.7 \cr
1.00   & 2.77 & 3.97  &  5.4  & 2.6--7.0 \cr
1.50   & 3.17 & 3.70  &  6.1  & 2.9--7.9 \cr
2.00   & 3.49 & 3.50  &  6.7  & 3.1--8.7 \cr
3.00   & 3.99 & 3.23  &  7.6  & 3.5--9.8 \cr  
\enddata
\end{deluxetable}

%\clearpage

\begin{deluxetable}{cccc}
%\tabletypesize{\scriptsize}
\tablecaption{Total Energy and Mass Transfer Rate\label{tabenergy}}
\tablewidth{0pt}
\tablehead{
\colhead{Assumed Distance} & \colhead{Assumed Recurrence}   &
\colhead{2-100 keV Fluence}  &  \colhead{Average Mass Transfer} \\
\colhead{(kpc)} & \colhead{Time (years)} &  
\colhead{(erg)}     &   \colhead{Rate ($M_{\odot}$ yr$^{-1}$)}
}
\startdata
3   &  10  &   $1.1\times 10^{45}$  &  $6.0\times 10^{-10}$ \cr
3   &  50  &   $1.1\times 10^{45}$  &  $1.2\times 10^{-10}$ \cr
6   &  10  &   $4.3\times 10^{45}$  &  $2.4\times 10^{-9}$ \cr
6   &  50  &   $4.3\times 10^{45}$  &  $4.8\times 10^{-10}$ \cr
10   &  10  &   $1.2\times 10^{46}$  &  $6.7\times 10^{-9}$ \cr
10   &  50  &   $1.2\times 10^{46}$  &  $1.3\times 10^{-9}$ \cr
\enddata
\end{deluxetable}


\begin{thebibliography}{}
\bibitem[Bailyn et al.(1995)]{bai95}
    Bailyn, C. D., Orosz, J. A., McClintock, J. E., \& Remillard,
    R. A. 1995, Nature, 378, 157
\bibitem[Bailyn et al.(1998)]{bai98}
    Bailyn, C. D., Jain, R. K., Coppi, P., \& Orosz, J. A. 1998, \apj,
    499, 367
\bibitem[Bailyn et al.(1999)]{bai99}
  Bailyn, C. D., 
  Depoy, D.,
  Agostinho, R., Mendez, R., Espinoza, J., \& Gonzalez, D. 1999, 
  American Astronomical Society Meeting 195, \#87.06
\bibitem[Bobinger(2000)]{bob00}
    Bobinger, A 2000, \aap, 357, 1170
\bibitem[Campbell-Wilson et al.(1998)]{cam98}
    Campbell-Wilson, D., McIntyre, V.,  Hunstead, R., \& Green, A.
    1998, IAU Circ.\ \#7010
\bibitem[Casares, Charles, \& Naylor(1992)]{cas92}
    Casares, J., Charles, P. A., \& Naylor, T. 1992, Nature, 355, 614
\bibitem[Casares et al.(1993)]{cas93}
    Casares, J., Charles, P. A., Naylor, T., \& Pavlenko, E. P. 1993,
    \mnras, 265, 834
\bibitem[Casares, Charles, \& Marsh(1995)]{cas95}
    Casares, J., Charles, P. A., \& Marsh, T. R. 1995, MNRAS, 277, L45
\bibitem[Charbonneau(1995)]{cha95}
    Charbonneau, P. 1995, \apjs, 101, 309
\bibitem[Chitre \& Hartle(1976)]{chi76}
    Chitre, D. M., \& Hartle, J. B. 1976, \apj, 207, 592
\bibitem[Cardelli, Clayton, \& Mathis(1989)]{car89}
    Cardelli, J. A., Clayton, G. C., Mathis, J. S. 1989, \apj,
    345, 245
\bibitem[Chen, Shrader, \& Livio(1997)]{che97}
    Chen, W., Shrader, C. R., \& Livio, M. 1997, \apj, 491, 312
\bibitem[Claret(2000)]{cla00}
    Claret, A. 2000, \aap, 359, 289
\bibitem[Cui et al.(1999)]{cui99}
    Cui, W., Zhang, S. N., Chen, W., \& Morgan, E. H. 1999, \apj, 512, L43
\bibitem[Echevarr\'ia et al.(1989)]{ech89}
    Echevarr\'ia, J., Diego, F., Tapia, M., Costero, R., Ruiz, E.,
    Salas, L., Guti\'erez, L., \& Enriquez, R., 1989, \mnras,
    240, 975
\bibitem[Regos, Tout, \& Wickramasinghe(1998)]{reg98}
    Regos, E., Tout, C., \& Wickramasinghe, D. 1998, \apj, 509, 362
\bibitem[Fich, Blitz, \& Stark(1989)]{fic89}
Fich, M., Blitz, L., \& Stark, A. A 1989, \apj, 342, 227
\bibitem[Filippenko, Matheson, \& Ho(1995)]{fil95a}
    Filippenko, A. V., Matheson, T., \& Ho, L. C. 1995, ApJ, 455, 614
\bibitem[Filippenko, Matheson, \& Barth(1995)]{fil95b}
    Filippenko, A. V., Matheson, T., \& Barth, A. J. 1995, \apj, 455, L139
\bibitem[Filippenko et al.(1999)]{fil99}
    Filippenko, A. V., Leonard, D. C.,
    Matheson, T., Li, W., Moran, E. C., \&
    Riess, A. G. 1999, PASP, 111, 696
\bibitem[Filippenko \& Chornock(2001)]{fil01}
    Filippenko, A. V., \& Chornock, R. 2001, IAU Circ.\ \#7644
\bibitem[Fryer \& Kalogera(2001)]{fry01}
    Fryer, C. L., \& Kalogera, V. 2001, \apj, 554, 548
\bibitem[Garcia et al.(2001)]{gar01}
    Garcia, M. R., McClintock, J. E., Narayan, R.,
    Callanan, P., Barret, D., Murray, S. S. 2001, \apj, 553, L47
\bibitem[Gotthelf et al.(1992)]{got92}
    Gotthelf, E., Halpern, J. P.,
    Patterson, J., \& Rich, R. M. 1992, \aj, 103, 219
\bibitem[Gray(1992)]{gra92}
    Gray, D. F. 1992, The Observation and Analysis of Stellar Photospheres,
    Cambridge Astrophys.\ Ser.\ 20 (Cambridge: Cambridge Univ.\ Press)
\bibitem[Greene, Bailyn, \& Orosz(2001)]{green01}
    Greene, J., Bailyn, C. D., \& Orosz, J. A. 2001, \apj, 554, 1290
\bibitem[Greiner, Cuby, \& McCaughrean(2001)]{gre01}
    Greiner, J., Cuby, J. G., \& McCaughrean, M. J. 2001, Nature, 
     414, 522
\bibitem[Hameury et al.(1997)]{ham97}
 Hameury, J.-M., Lasota, J.-P.,
 McClintock, J. E., \& Narayan, R. 1997, \apj, 489, 234
\bibitem[Hannikainen et al.(2001)]{han01b}
   Hannikainen, D., 
   Campbell-Wilson, D., 
   Hunstead, R., McIntyre, V., 
   Lovell, J., Reynolds, J.,  Tzioumis, T., \& Wu, K. 2001, in
   Proceedings of the Third Microquasar Workshop: 
   Granada Workshop on galactic relativistic jet sources, Eds.\ A. J.
   Castro-Tirado, J. Greiner \& J. M. Paredes, 
   Ap\&SS, 276, 45  
\bibitem[Harlaftis et al.(1997)]{har97}
    Harlaftis, E. T., Steeghs, D.,
    Horne, K., \& Filippenko, A. V. 1997, \aj, 114, 117
\bibitem[Hauschildt, Allard, \& Baron(1999a)]{hau99a} Hauschildt, P. H., 
    Allard, F., \& Baron, E. 1999a, \apj, 512, 377
\bibitem[Hauschildt et al.(1999b)]{hau99b} Hauschildt, P. H., Allard, F.,
    Ferguson, J., Baron, E., \& Alexander, D. R. 1999b, \apj, 525, 871
\bibitem[Homan et al.(2001)]{hom01}
    Homan, J., Wijnands, R., van der Klis, M., Belloni, T., van Paradijs, J.,
    Klein-Wolt, M., Fender, R., \& M\'endez, M. 2001, \apjs, 132, 377
\bibitem[Horne \& Marsh(1986)]{hor86}
    Horne, K., \& Marsh, T. R. 1986, MNRAS, 218, 761
\bibitem[Horne, Wade, \& Szkody(1986)]{hor86a}
    Horne, K., Wade, R. A., \& Szkody, P. 1986, \mnras, 219, 791
\bibitem[Jain et al.(1999)]{jai99}
    Jain, R. K., Bailyn, C. D., Orosz, J. A., Remillard, R. A., \& McClintock,
    J. E. 1999, \apj, 517, L131
\bibitem[Jain, Bailyn, \& Tomsick(2001a)]{jai01a}
    Jain, R., Bailyn, C., \& Tomsick, J. 2001a, IAU Circ.\ \# 7575
\bibitem[Jain et al.(2001b)]{jai01b}
    Jain, R. K., Bailyn, C. D., Orosz, J. A.,
    McClintock, J. E., Sobczak, G. J., \& 
    Remillard, R. A. 2001b, \apj, 546, 1086
\bibitem[Jain et al.(2001c)]{jai01c}
    Jain, R. K., Bailyn, C. D.,
    Orosz, J. A., McClintock, J. E., \&
    Remillard, R. A. 2001c, \apj, 554, L181
\bibitem[Johnston, Kulkarni, \& Oke(1989)]{joh89}
    Johnston, H. M., Kulkarni, S. R., \& Oke, J. B. 1989, \apj, 345, 492
\bibitem[Kalogera \& Baym(1996)]{kal96}
    Kalogera, V., \& Baym, G. 1996, \apj, 470, L61
\bibitem[Kolb et al.(1997)]{kol97}
    Kolb, U., King, A. R., Ritter, H., \& Frank, J. 1997, \apj, 485, L33
\bibitem[Marsh, Robinson, \& Wood(1994)]{mar94}
    Marsh, T. R., Robinson, E. L., \& Wood, J. H. 1994, \mnras, 266, 137
\bibitem[McClintock \& Remillard(1986)]{mcc86}
    McClintock, J. E., \& Remillard, R. A. 1986, \apj, 308, 110
\bibitem[McClintock et al.(2001)]{mcc01}
     McClintock, J. E., Garcia, M. R., Caldwell, N., Falco, E. E.,
     Garnavich, P. M., \&  Zhao, P. 2001, \apj, 511, L147
\bibitem[Metcalfe(2001)]{met01}
Metcalfe, T. S. 2001, PhD. Thesis, University of Texas, Austin
(http://whitedwarf.org/metcalfe)
\bibitem[Mirabel \& Rodriguez(1999)]{mir99}
    Mirabel, I. F., \& Rodriquez, L. F. 1999, \araa, 37, 409
\bibitem[Narayan, McClintock, \& Yi(1996)]{nar96}
    Narayan, R., McClintock, J. E., \& Yi, I. 1996, \apj, 457, 821
\bibitem[Orosz et al.(1994)]{oro94}
    Orosz, J. A., Bailyn, C. D., Remillard, R. A.,  McClintock, J. E.,
    \& Foltz, C. B. 1994, \apj, 436, 848
\bibitem[Orosz \& Bailyn(1995)]{oro95}
    Orosz, J. A., \& Bailyn, C. D. 1995, \apj, 446, L59
\bibitem[Orosz \& Bailyn(1997)]{oro97}
    Orosz, J. A., \& Bailyn, C. D. 1997, \apj, 477, 876
\bibitem[Orosz et al.(1997)]{oro97a}
    Orosz, J. A., Remillard, R. A., Bailyn, C. D., \& McClintock, J. E.
    1997, \apj, 478, L83
\bibitem[Orosz et al.(1998)]{oro98}
    Orosz, J. A., Jain, R. K., Bailyn, C. D., McClintock, J. E., \& 
    Remillard, R. A.
    1998, \apj, 499, 375
\bibitem[Orosz, Jain, \& Bailyn(1998)]{oro98a}
    Orosz, J. A., Jain, R. K., \& Bailyn, C. D. 1998, IAU Circ.\ \#7009
\bibitem[Orosz \& Hauschildt(2000)]{oro00}
    Orosz, J. A., \& Hauschildt, P. H. 2000, \aap, 364, 265
\bibitem[Orosz et al.(2001)]{oro01}
    Orosz, J. A., Kuulkers, E., van der Klis, M., McClintock, J. E.,
    Garcia, M. R., Callanan, P. C., Bailyn, C. D., Jain, R. K., \& Remillard,
    R. A. 2001, \apj, 555, 489
\bibitem[Predehl \& Schmitt(1995)]{pre95}
    Predehl, P., \& Schmitt, J. H. M. M. 1995, \aap, 293, 889
\bibitem[Remillard, McClintock, \& Bailyn(1992)]{rem92}
    Remillard, R. A., McClintock, J. E., \& Bailyn, C. D. 1992, \apj, 399,
    L145
\bibitem[Remillard et al.(1996)]{rem96}
    Remillard, R. A., Orosz, J. A.,
    McClintock, J. E., Bailyn, C. D. 1996, \apj, 459, 226
\bibitem[Remillard et al.(1999)]{rem99}
    Remillard, R. A., McClintock, J. E., 
    Sobczak, G. J., Bailyn, R. A., Orosz, J. A., Morgan, E. H., 
    \& Levine, A. M.
    1999, \apj, 517, L127
\bibitem[Remillard(2001)]{rem01}
    Remillard, R. A. 2001, 
    in Evolution of Binary and Multiple Stars, eds.\ 
     P. Podsiadlowski, S. Rappaport, A. King, F.
     D'Antona, and L. Burderi (San Francisco: ASP), in press (astro-ph/0103431)
\bibitem[Remillard et al.(2002)]{rem01b}
    Remillard, R. A., Muno, M. P., Sobczak, G. J., \& McClintock, J. E.
    2002, \apj, in press (vol.\ 465)
\bibitem[Shapiro \& Teukolsky(1983)]{sha83}
    Shapiro, S. L., \& Teukolsky, S. L. 1983, 
    Black Holes, White Dwarfs, and Neutron Stars (New York: Wiley) 
\bibitem[Sobczak et al.(1999a)]{sob99}
    Sobczak, G. J., McClintock, J. E.,
    Remillard, R. A., Levine, A. M., Morgan, E. H.,
    Bailyn, C. D., \& Orosz, J. A. 1999a, \apj, 517, L121
\bibitem[Sobczak et al.(1999b)]{sob99b}
    Sobczak, G. J., McClintock, J. E., Remillard, R. A., Bailyn, C. D.,
    \& Orosz, J. A. 1999b, \apj, 520, 776
\bibitem[Sobczak et al.(2000)]{sob00}
    Sobczak, G. J., McClintock, J. E., Remillard, R. A., Cui, W., Levine,
    A. M., Morgan, E. H., Orosz, J. A., \& Bailyn, C. D. 2000, \apj, 544,
    993
\bibitem[Smak(1981)]{sma81}
    Smak, J. 1981, Acta Astron., 31, 395
\bibitem[Shahbaz et al.(1994)]{sha94}
    Shahbaz, T., Ringwald, F. A., Bunn, J. C., Naylor, T., Charles, P. A.,
    \& Casares, J. 1994, \mnras, 271, L10
\bibitem[Smith et al.(1998)]{smi98}
    Smith, D. A., et al.\ 1998, IAU Circ.\ 
    \#7008
\bibitem[Stetson(1987)]{ste87} 
   Stetson, P. B., 1987, PASP, 99, 191
\bibitem[Stetson(1990)]{ste90}
   Stetson, P. B., 1990, PASP, 102,  932
\bibitem[Stetson, Davis, \& Crabtree(1991)]{ste91}
  Stetson, P. B., Davis, L. E., Crabtree, D. R., 1991, in
  ``CCDs in Astronomy,'' ed.\ G. Jacoby, ASP Conference Series, Volume
  8, page 282
\bibitem[Stetson(1992a)]{ste92a} 
  Stetson P. B., 1992a, in ``Astronomical Data Analysis Software and 
  Systems I,'' eds.\ D. M. Worrall, C. Biemesderfer, \& J. Barnes, ASP 
  Conference Series, Volume 25, page 297
\bibitem[Stetson(1992b)]{ste92b} 
  Stetson, P. B., 1992b, in ``Stellar Photometry--Current Techniques and 
  Future Developments,'' IAU Coll.\ 136, eds.\ C. J. Butler, \& I. Elliot, 
  Cambridge University Press, Cambridge, England,
  page 291
\bibitem[Strai\v{z}ys \& Kuriliene(1981)]{str81}
    Strai\v{z}ys, V., \& Kuriliene, G. 1981, Ap\&SS, 80, 353
\bibitem[Strohmayer(2001)]{str01}
    Strohmayer, T. 2001, \apj, 552, L49
\bibitem[Tonry \& Davis(1979)]{ton79}
    Tonry, J., \& Davis, M. 1979, \aj, 84 1511
\bibitem[van Paradijs(1996)]{van96}
   van Paradijs, J. 1996, \apj, 464, L139
\bibitem[Vtrilek et al.(1991)]{vrt91}
Vrtilek, S. D., McClintock, J. E., Seward, F. D., 
Kahn, S. M., \& Wargelin, B. J. 1991, \apjs, 76, 1127
\end{thebibliography}
\end{document}